\documentclass[12pt]{iopart}

\usepackage{amssymb}
\usepackage{amsbsy}
\usepackage[dvips]{graphicx}
\usepackage{color}

\newcommand{\be}{\begin{equation}} \newcommand{\ee}{\end{equation}}
\newcommand{\bea}{\begin{eqnarray}} \newcommand{\eea}{\end{eqnarray}}
\newcommand{\el}{\nonumber \\}
\newcommand{\re}[1]{(\ref{#1})}
\newcommand{\pat}{\partial}
\newcommand{\adot}{\dot{a}}

\newcommand{\sigmadot}{\dot{\sigma}}
\newcommand{\sigmaddot}{\ddot{\sigma}}
\newcommand{\rhodot}{\dot{\rho}}
\newcommand{\Hdot}{\dot{H}}

\newcommand{\bk}{\bi{k}}
\newcommand{\brt}[1]{[#1]}
\newcommand{\kt}{\tilde{k}}
\newcommand{\sR}{\mathcal{R}}
\newcommand{\sS}{\mathcal{S}}

\newcommand{\hsR}{\hat{\mathcal{R}}}
\newcommand{\hsS}{\hat{\mathcal{S}}}

\newcommand{\ahat}{\hat{a}}

\newcommand{\PRD}[1]{{\it Phys. Rev.} {\bf D#1}}
\renewcommand{\PRL}[1]{Phys. Rev. Lett. {\bf #1}}
\newcommand{\NPA}[1]{{\it Nucl. Phys.} {\bf A#1}}
\newcommand{\NPB}[1]{{\it Nucl. Phys.} {\bf B#1}}
\newcommand{\PLB}[1]{{\it Phys. Lett.} {\bf B#1}}
\newcommand{\MNRAS}[1]{{\it Mon. Not. Roy. Astron. Soc.} {\bf #1}}
\newcommand{\APJ}[1]{{\it Astrophys. J.} {\bf #1}}

\newcommand{\m}[3]{${#1}_{- #2}^{+ #3}$}

\begin{document}

\begin{flushleft}
	\hfill			HIP-2004-30/TH \\ \hfill \\
\end{flushleft}

\title{Correlated isocurvature perturbations from mixed inflaton-curvaton decay}

\author{Francesc Ferrer\dag, Syksy R\"{a}s\"{a}nen\ddag \ and Jussi V\"{a}liviita$^\ast$}

\address{\dag Astrophysics, University of Oxford,
Denys Wilkinson Building, Keble Road, Oxford, OX1 3RH, UK}

\address{\ddag Theoretical Physics, University of Oxford,
1 Keble Road, Oxford, OX1 3NP, UK}

\address{$^\ast$Department of Physical Sciences, University of Helsinki and Helsinki Institute of Physics, PO Box 64, FIN-00014 University of Helsinki, Finland}

\eads{\mailto{f.ferrer1@physics.ox.ac.uk}, \mailto{syksy.rasanen@iki.fi}, \mailto{jussi.valiviita@helsinki.fi}}

\begin{abstract}

\noindent
We study cosmological perturbations in the case that
present-day matter consists of a mixture of inflaton
and curvaton decay products.
We calculate how the curvaton perturbations are transferred
to its decay products in the general case when it does
not behave like dust.
Taking into account that the decay products of the inflaton
can also have perturbations results in an interesting
mixture of correlated adiabatic and isocurvature perturbations.
In particular, negative correlation can
improve the fit to the CMB data by lowering the angular power
in the Sachs-Wolfe plateau without changing the peak structure.
We do an 11-parameter fit to the WMAP data. We find that
the best-fit is not the 'concordance model', and that
well-fitting models do not cluster around the best-fit,
so that cosmological parameters cannot be reliably
estimated. We also find that in our model the mean
quadrupole ($l=2$) power is $l(l+1) C_l /2\pi = 1081 \mu$K$^2$,
much lower than in the pure adiabatic $\Lambda$CDM model, which
gives $1262 \mu$K$^2$.

\end{abstract}



\tableofcontents

\setcounter{secnumdepth}{3}

\section{Introduction}

The building blocks of what is today known as the curvaton
model were introduced in \cite{Mollerach:1990, Linde:1996}
(some early discussion can be found in \cite{Linde, Kofman:1986}).
The first detailed application
to the observed CMB anisotropies was done in \cite{Enqvist:2001},
in the context of the pre-big bang scenario. The model
was then applied to inflation in detail and given the name
``curvaton'' in \cite{Lyth:2001, Moroi:2001}.

In the curvaton model the perturbations of the cosmic
microwave background and large-scale structure originate
from a field, the curvaton, which is different from the
field(s) driving the dynamics of the very early universe,
be it inflation, pre-big bang collapse or something else.
In what follows, we adopt the language of
single field inflation for describing the primordial universe,
though our considerations could apply to other cosmological
scenarios as well.

During inflation and reheating, the curvaton is assumed to be
completely subdominant. The mass of the curvaton field is also
assumed to be much less than the Hubble parameter during inflation,
so that the curvaton field value is frozen, and the curvaton
will acquire a spectrum of quantum fluctuations which
become classical. After inflation, when the Hubble parameter
falls and becomes of the order of the curvaton mass, the curvaton
will start to oscillate around the minimum
of its potential. Assuming the potential to be quadratic
about the minimum, the energy density in the curvaton field,
averaged over oscillations, decays like $a^{-3}$,
where $a$ is the scale factor. Taking the universe after reheating
to be dominated by radiation, the relative contribution
of the curvaton to the energy density increases. (Since
the curvaton field value and therefore energy density is constant
before the start of the oscillations, its relative contribution
increases even faster during that period.)

We assume that the curvaton decays into baryons, cold dark matter (CDM),
leptons and photons. The leptons are assumed to consist of
(or later decay into, depending on when the curvaton decays)
electrons and neutrinos.
(Electrons play no role in our discussion, and will not be
mentioned from now on.) If the curvaton decays before
it contributes to the background, the perturbations
inherited by its decay products will be of the isocurvature
type, whereas if the curvaton completely determines the background when
it decays, the perturbations will be adiabatic.
This conversion of isocurvature perturbations into adiabatic
ones is the essence of the curvaton mechanism.
If the curvaton decays between these two extremes, its decay
products will inherit a mixture of correlated adiabatic and
isocurvature perturbations \cite{Gordon:2000hv}. This has been studied in
\cite{Moroi:2002, Lyth:2002, Sloth:2002, Malik:2002, Gordon:2002, Lyth:2003, Gordon:2003, Gupta:2003, Langlois:2004}.

In the present paper, we extend in two respects
previous studies (building particularly on \cite{Gupta:2003})
of the case that the curvaton decays while it contributes to
the energy density but does not completely dominate it.

First, it has usually been assumed that one can treat the
curvaton as a pressureless fluid in calculating its decay.
However, as this description is valid only as an average over
oscillations, one would not expect it to apply
in the case of a rapid decay, or when the field
decays just as it is starting to oscillate (recently studied
in \cite{Langlois:2004}). We will calculate the decay without
this assumption, using the curvaton field equation, and will
obtain the limit of validity of the dust approximation.

Second, it has usually been assumed that even when present-day
radiation and matter consist of a mixture of inflaton and
curvaton decay products, the inflaton decay products have no
perturbations. We will take into account the possibility
that both the inflaton and curvaton decay products carry some
(uncorrelated) perturbations, resulting in an interesting mixture.
This case has been studied in
\cite{Moroi:2001, Moroi:2002, Langlois:2004}
and is also reminiscent of double inflation models
\cite{Linde, Kofman:1987, Kofman:1988, Polarski:1994, Langlois:1999a, Tsujikawa:2002}.
Two limiting cases are the
usual inflaton scenario, where all present-day matter comes from the
inflaton, and the original pre-big bang curvaton scenario
\cite{Enqvist:2001}, where everything comes from the curvaton.

In section \ref{decay} we consider the decay of the curvaton,
solve the decay equations numerically and discuss recovering
the dust approximation. In section \ref{WMAP} we fit our spectrum
of perturbations to the measurements made by the WMAP
satellite. We discuss the interesting features of the spectrum,
in particular suppression of the low multipoles and negative
running of the spectral index. Having two independent
sources of perturbations allows for a spectrum where one
can have different behaviour of the low and high multipoles.
The acoustic peak structure can remain very close
to the pure adiabatic case, while the angular power
on the Sachs-Wolfe plateau can be much reduced, leading to
better overall fit than the pure adiabatic $\Lambda$CDM model.
We also find that the inclusion of isocurvature modes
opens up the space of well-fitting models,
so that one cannot determine the cosmological
parameters from the WMAP data alone.
A detailed study of the estimation of the isocurvature contribution
and the cosmological parameters with our perturbation spectrum will
be presented later \cite{Ferrer:2004}.

\section{The curvaton decay} \label{decay}

\subsection{The set-up}

We assume that there is a bath of baryons, CDM,
neutrinos and photons produced by the decay of the inflaton field.
We take the perturbations in these fluids to be adiabatic,
as expected from single field models of inflation \cite{Weinberg:2004a}.
We also assume that there is a curvaton field which is
subdominant, but frozen at some field value so that its
relative contribution to the energy density rises. The curvaton
field is assumed to have a spectrum of perturbations,
acquired during inflation, which is uncorrelated with that
of the inflaton decay products.

Like the inflaton, the curvaton is assumed to decay into
baryons, CDM, neutrinos and photons,
which inherit the perturbations of the curvaton.
The final state will therefore consist of a mixture of
products of inflaton and curvaton decay, both with their
own spectrum of perturbations.

We will follow the system numerically from the time that the
curvaton is frozen and subdominant until the time it has completely
decayed. We will see how the resulting spectrum of perturbations
in the decay products depends on the initial conditions
and parameters of the model. Our analysis of the decay is a
generalisation of that in \cite{Gupta:2003}, where the curvaton
decay was followed numerically assuming that the curvaton behaves
as a pressureless fluid.

\subsection{The equations} \label{equations}

\paragraph{The background.}

The spacetime is a perturbed Friedmann-Robertson-Walker universe.
In accordance with the generic inflationary prediction, we take
the background spacetime to be spatially flat. In the uniform
curvature gauge, the metric reads (we consider only scalar perturbations)
\bea \label{metric}
  \rmd s^2 = -(1+2\phi) \rmd t^2 + 2 a B_{,i}\rmd t \rmd x^i + a^2 \delta_{ij} \rmd x^i \rmd x^j \ .
\eea

Throughout, we follow the notational conventions of
\cite{Malik:2002, Gupta:2003, Malik:2001}, except that we prefer
to use $\sR$, the curvature perturbation on the comoving hypersurface,
rather than $\zeta$, the curvature perturbation on the uniform
energy density hypersurface. On superhorizon scales, 
the two are related by $\sR=-\zeta$.

As sources, we have radiation ($r$), matter ($m$)\footnote{The identification
of radiation and matter as baryons, CDM, neutrinos and photons will be
discussed in section \ref{spectrum}.} and the curvaton field ($\sigma$).
The relation between the background sources and geometry is given by
\bea 
  \label{Hubble} H^2 = \frac{1}{3 M_{Pl}^2}\rho \\
  \label{backcons} \rhodot + 3 H ( \rho + P ) = 0 \ ,
\eea

\noindent where $H=\adot/a$, $M_{Pl}=1/\sqrt{8 \pi G_{\textrm{N}}}$ is
the (reduced) Planck mass and $\rho$ and $P$ are the total energy density
and pressure, respectively. In terms of the individual components, we
have
\bea
  \Sigma_{\alpha} \rho_{\alpha} = \rho \ , \quad \Sigma_{\alpha} P_{\alpha} = P \ ,
\eea

\noindent where $\alpha=r, m, \sigma$. For radiation and
matter we have $P_{r}=\frac{1}{3}\rho_{r}, P_{m}=0$, so that
if there was no energy transfer between the components, their
energy densities would decay like
$\rho_{r}\propto a^{-4}$ and $\rho_{m}\propto a^{-3}$. Since
the curvaton decay transfers energy into radiation and matter,
the behaviour of the energy densities is given instead by
\bea \label{backdens}
  \rhodot_{\alpha} + 3 H ( \rho_{\alpha} + P_{\alpha} ) = Q_{\alpha} \ ,
\eea

\noindent where $Q_{\alpha}$ is the energy density transfer
per unit time to the fluid $\alpha$.

We take the curvaton field to be minimally coupled to gravity and to
have the potential $V(\sigma)=\frac{1}{2} m^2 \sigma^2$. We model the
curvaton decay phenomenologically by introducing constant decay
rates to both radiation and matter into the equation of
motion, as in \cite{Gupta:2003}. The curvaton equation of
motion for the background then reads
\bea \label{backeom}
  \sigmaddot + (3 H + \Gamma) \sigmadot + m^2 \sigma = 0 \ ,
\eea

\noindent where $\sigma$ is the background value of the curvaton
field and $\Gamma=\Gamma_r+\Gamma_m$ is the
curvaton decay rate, with $\Gamma_r$ and $\Gamma_m$ being
constants. Using this effective description of decay, we ignore
the possibility of parametric resonance
\cite{Traschen:1990, Kofman:1997}, as is usually done for
curvaton models. Thermal corrections to the potential
and to the decay rate \cite{Enqvist:2004pr} could also be important,
since the curvaton is coupled to the thermal bath of matter and
radiation \cite{Postma:2002}; we ignore these too.

The energy density and pressure of the curvaton are given by
\bea \label{backsigma}
  \rho_{\sigma} = \frac{1}{2} \sigmadot^2 + \frac{1}{2} m^2 \sigma^2 \el
  P_{\sigma} = \frac{1}{2} \sigmadot^2 - \frac{1}{2} m^2 \sigma^2 \ ,
\eea

\noindent and the energy transfers are
\bea \label{backtrans}
  Q_r = \Gamma_r (\rho_{\sigma} + P_{\sigma}) \el 
  Q_m = \Gamma_m (\rho_{\sigma} + P_{\sigma}) \el 
  Q_{\sigma} = - (\Gamma_r+\Gamma_m) (\rho_{\sigma} + P_{\sigma}) \ .
\eea

\noindent Note that $\rho_{\sigma} + P_{\sigma}=\sigmadot^2$.
In \cite{Gupta:2003}, it was noted that we have an upper limit
$\Gamma_m/\Gamma_r\lesssim 10^{-6}$ from the requirement that
the curvaton decay should be complete before
big bang nucleosynthesis (BBN) at $z\sim 10^{10}$
and matter should be subdominant until the matter-radiation equality at
$z\lesssim 10^4$. This limit only holds if all matter and
radiation come from the curvaton, which is not true in
our case (or in the case of \cite{Gupta:2003}). We will
discuss the limit on $\Gamma_m/\Gamma_r$ in section
\ref{perturbations}.

For comparison with the dust treatment of the curvaton in
\cite{Gupta:2003}, let us define \cite{Turner:1983}:
\bea \label{phisq}
  \rho_{\sigma} + P_{\sigma} \equiv \left(\gamma+\gamma_p \right) \rho_{\sigma} \ ,
\eea

\noindent where $\gamma\rho_{\sigma}$ and $\gamma_p\rho_{\sigma}$
are the average and the periodic part, respectively, of
$\rho_{\sigma} + P_{\sigma}$ over an oscillation.
In the regime where the curvaton oscillation is much
more rapid than the decay, $m\gg\Gamma$, we can neglect
$\gamma_p$. Moreover, for a quadratic potential
$\gamma=1$, so that in the rapid oscillation limit we
approximately recover the expressions of \cite{Gupta:2003}:
\bea \label{qapprox}
  Q_r = \Gamma_r \rho_{\sigma} \el
  Q_m = \Gamma_m \rho_{\sigma} \ .
\eea 

\noindent However, we do not expect \re{qapprox} to be a good
approximation when the field does not oscillate rapidly when it decays.

\paragraph{The perturbations.}

We are interested in the behaviour of perturbations only
in the large-scale limit, when spatial gradients can be neglected.
Then, the equations for the perturbations read (after eliminating
the metric perturbation $\phi$) \cite{Malik:2002, Gupta:2003}
\bea \label{pertdens}
  \delta\rhodot_{\alpha} + 3 H ( \delta\rho_{\alpha} + \delta P_{\alpha} ) = \delta Q_{\alpha} - \frac{\delta\rho}{2\rho} Q_{\alpha} \ ,
\eea

\noindent where $\delta\rho_{\alpha}, \delta P_{\alpha}$ and
$\delta Q_{\alpha}$ are, respectively, the perturbation in the
energy density and pressure of, and energy transfer to, the
component $\alpha$. The total perturbation obeys the covariant
conservation equation
\bea
  \label{pertcons} \delta\rhodot + 3 H ( \delta\rho + \delta P ) = 0 \ ,
\eea

\noindent where
\bea
  \delta\rho = \Sigma_{\alpha} \delta\rho_{\alpha} \ , \quad \delta P = \Sigma_{\alpha}\delta P_{\alpha} \ .
\eea

The equation of motion for the curvaton field perturbation $\delta\sigma$ is
\bea \label{perteom}
  \fl \delta\sigmaddot + (3 H + \Gamma) \delta\sigmadot + m^2 \delta\sigma = \left( 2 m^2\sigma + \Gamma\sigmadot \right) \frac{\delta\rho}{2\rho} + 3 H\sigmadot\left( \frac{\delta P}{\delta\rho} - \frac{P}{\rho} \right) \frac{\delta\rho}{2\rho} \ .
\eea

\noindent The perturbations in the energy density and pressure of the
curvaton are given by
\bea \label{pertsigma}
  \delta\rho_{\sigma} = \sigmadot\delta\sigmadot + m^2 \sigma\delta\sigma + \frac{\delta\rho}{2\rho} \sigmadot^2 \el
  \delta P_{\sigma} = \sigmadot\delta\sigmadot - m^2 \sigma\delta\sigma + \frac{\delta\rho}{2\rho} \sigmadot^2 \ ,
\eea

\noindent and the perturbed energy transfers are
\bea \label{perttrans}
  \delta Q_{r} = \Gamma_r (\delta\rho_{\sigma} + \delta P_{\sigma}) \el 
  \delta Q_{m} = \Gamma_m (\delta\rho_{\sigma} + \delta P_{\sigma}) \el 
  \delta Q_{\sigma} = - (\Gamma_r+\Gamma_m) (\delta\rho_{\sigma} + \delta P_{\sigma}) \ .
\eea

\noindent We have assumed that the decay rates are constant
also at the perturbed level, as in \cite{Gupta:2003}.

\subsection{Behaviour of the background} \label{background}

The parameters of the background are the curvaton mass $m$,
the decay rate $\Gamma$, the ratio $\Gamma_m/\Gamma_r$,
as well as the initial values
which consist of the initial curvaton field value $\sigma_0$
and the initial radiation energy density $\rho_{r0}$.
Under the assumption that matter is subdominant,
the initial energy density of matter makes no difference.
Since we assume that initially $H\gg m,\Gamma$, we see from
\re{backeom} that the curvaton is
essentially frozen at a constant value to which it
has been driven during inflation, so its initial
velocity is negligible (though non-zero). For the curvaton
to be subdominant, we must have $\rho_{r0}\gg\frac{1}{2} m^2 \sigma_0^2$. 

When the Hubble parameter becomes of the order of the curvaton
mass, $H\sim m$, the curvaton rolls down the potential and
starts to oscillate.
When the Hubble parameter becomes of the order of the decay rate,
$H\sim\Gamma$, the curvaton decays into radiation and matter. However,
since the energy transfer in \re{backtrans} is proportional to
$\sigmadot^2$, there is no decay as long as the field is
frozen. This means that if $m<\Gamma$, the decay will not start at
$H\sim\Gamma$, but only later at $H\sim m$, when the field will
simultaneously roll down the potential and decay. So, the condition
for curvaton decay is $H\sim$ min($m,\Gamma$). Note that this makes
it a natural possibility (in the context of our phenomenological
treatment) for the curvaton to decay already when it starts
to oscillate.

If $\sigma_0\gtrsim M_{Pl}$, we see from \re{Hubble} and
\re{backdens} that the curvaton starts contributing significantly to
the energy density before it starts to oscillate.
For $\sigma_0 \gg M_{Pl}$, a period of curvaton-driven inflation will ensue
and the inflaton-curvaton model will in fact be a double inflation
model. In a realistic model, the curvaton
potential is not expected to be exactly quadratic due to Planck scale
-suppressed non-renormalisable terms, so we should not apply our
calculation to the case $\sigma_0\gg M_{Pl}$.
At any rate, we want to study the case when both the
inflaton and curvaton decay products contribute to the observed
radiation and matter, so the region of parameter space where there is a long
second period of inflation is of no interest to us.
Note that if in Planck-scale inflation the values of fields such
as the curvaton randomly sample a more-or-less uniform distribution,
a value $\sigma_0\sim M_{Pl}$ is much more likely than a value
$\sigma_0\ll M_{Pl}$. For values $\sigma_0\sim M_{Pl}$ and $\Gamma\gtrsim m$,
the curvaton will decay as it is starting to contribute to the
energy density, making it a not unnatural possibility that both
the inflaton and curvaton contribute to the radiation
and matter density.

\subsection{Behaviour of the perturbations} \label{perturbations}

In the beginning, the system consists of radiation and matter
from the inflaton decay, carrying one spectrum of perturbations, and
the curvaton, carrying another, uncorrelated, spectrum of
perturbations. In the final state after curvaton decay, only
radiation and matter, carrying  a mixture of the two spectra, are
left.

The curvature perturbation of the component $\alpha$ is (in
the uniform curvature gauge)
\bea \label{Ralpha}
  \sR_{\alpha} = H \frac{\delta\rho_{\alpha}}{\rhodot_{\alpha}} \ ,
\eea
and the total curvature perturbation is
\bea
  \sR  =  \Sigma_{\alpha} \frac{\rhodot_{\alpha}}{\rhodot}\sR_{\alpha} = H \frac{\delta\rho}{\rhodot} \ .
\eea

We are interested in how the perturbations of the curvaton are
transferred to its decay products. Usually, the curvaton is
assumed to behave like dust, $P_{\sigma}=0$. Since the equation
of state is then (trivially)
barotropic, $P_{\sigma}= P_{\sigma}(\rho_{\sigma})$,
the curvature perturbation $\sR_{\sigma}$
is conserved on superhorizon scales \cite{Wands:2000}.
Then the ratio $\sR_{\alpha}^{out}/\sR_{\sigma}^{in}$
(using only the part of $\sR_{\alpha}^{out}$ that comes
from the curvaton decay) gives a useful measure of how
the curvaton perturbations are transferred to the fluid $\alpha$.
When the curvaton is treated like dust we
have, from \re{backdens}, \re{backtrans} and \re{Ralpha},
\bea \label{Rdust}
  \sR_{\sigma}(P_{\sigma}=0) = - H\frac{\delta\rho_{\sigma}}{(3 H + \Gamma)\rho_{\sigma}} \ .
\eea

However, when we treat the curvaton as a field and not as dust,
it does not have a barotropic equation of state, as
is clear from \re{backsigma}, and as a result $\sR_{\sigma}$
is not conserved. In this case the curvaton perturbation
$\sR_{\sigma}$ is, from \re{backdens}, \re{backsigma},
\re{backtrans} and \re{Ralpha},
\bea \label{Rfield}
  \sR_{\sigma} = - H\frac{\delta\rho_{\sigma}}{(3 H + \Gamma)\sigmadot^2} \ .
\eea

Before the curvaton starts to roll down the potential,
the field is nearly frozen and $\sigmadot$ is small.
As the curvaton starts to oscillate, the absolute value
of $\sigmadot$ increases significantly, which
translates into a large decrease of $\sR_{\sigma}$.
(When the curvaton field oscillates, $\sR_{\sigma}$
oscillates around a fixed value, so treating it as
constant is then justified, when averaged over oscillations.)

Therefore, giving the initial curvaton perturbations
in terms of $\sR_{\sigma}^{in}$
and calculating the ratios $\sR_{\alpha}^{out}/\sR_{\sigma}^{in}$
is unenlightening, because the results depend strongly
on when the initial 
conditions are given. To meaningfully measure the transfer of the
curvaton perturbations to its decay products, it is better to
consider some conserved quantity. We therefore \emph{define}
a convenient perturbation variable $\xi_{\sigma}$ by saying that
it has the same form \re{Rdust} as in the dust case,
\bea \label{xisigma}
  \xi_{\sigma} \equiv - H \frac{\delta\rho_{\sigma}}{(3 H + \Gamma)\rho_{\sigma}} \ ,
\eea

\noindent but with $\rho_{\sigma}$ and $\delta\rho_{\sigma}$
given by the expressions for the field case,
\re{backsigma} and \re{pertsigma}. This quantity is
constant when the curvaton field is frozen and $H\gg\Gamma$
(as we assume initially to be the case). The initial value of
$\xi_{\sigma}$ is essentially the initial value of
$-\delta\rho_{\sigma}/(3\rho_{\sigma})$, so it is a useful 
measure of the curvaton perturbations.
We will denote the initial value of $\xi$ by
$\xi_{\sigma}^{in}=\sR_2$, and will be interested in the ratios
$\sR_{\alpha}^{out}/\sR_2$. The quantity $\sR_2$ is related to
the initial curvaton curvature perturbation by
\bea \label{Rxi}
  \sR_{\sigma}^{in} &=& \frac{\rho_{\sigma0}}{\sigmadot_0^2} \sR_2 \el
  &=& \frac{1}{2}\left( 1 + \frac{m^2\sigma_0^2}{\sigmadot_0^2} \right) \sR_2 \ , 
\eea
where the subscript $0$ refers to the initial value. We have
traded the strong dependence on $\sigmadot^2$ for a large
normalisation factor of the initial perturbation.

So, in the beginning we have the following perturbations
\bea \label{Rinit}
  \sR_r^{in} &=& \sR_1 \el
  \sR_m^{in} &=& \sR_1 \el
  \xi_{\sigma}^{in} &=& \sR_2 \ ,
\eea
where $\sR_1$ is the spectrum of perturbations
inherited by the inflaton decay products, which is
uncorrelated with $\sR_2$.
In the beginning, the universe is radiation-dominated, so the
total curvature perturbation is $\sR^{in}=\sR_r^{in}=\sR_1$.

After the end of curvaton decay we have
\bea
  \label{Rr} \sR_r^{out} &=& (1-A_r)\sR_1 + A_r\lambda_r\sR_2 \\
  \label{Rm} \sR_m^{out} &=& (1-A_m)\sR_1 + A_m\lambda_m\sR_2 \ .
\eea
As noted earlier, the curvaton decay has to be complete before
the onset of BBN, so the universe is still radiation-dominated. The
total curvature perturbation and the matter-radiation isocurvature
perturbation \mbox{$\sS_{m,r}^{out}=-3(\sR_m-\sR_r)$} are then
\bea
  \label{R} \sR^{out} &=& (1-A_r)\sR_1 + A_r\lambda_r\sR_2 \\
  \label{S} \sS_{m,r}^{out} &=& 3 (A_m-A_r)\sR_1 - 3 (A_m\lambda_m-A_r\lambda_r)\sR_2 \ .
\eea

In \re{Rr}, \re{Rm}, \re{R} and \re{S}, there are two
different kinds of factors relating the final
perturbations to the initial ones. First, the dilution factors
$A_{\alpha}$ give the percentage of the fluid $\alpha$ coming
from the curvaton,
\bea \label{dilution}
  A_{\alpha} = \frac{\rho_{\alpha2}}{ \rho_{\alpha1} + \rho_{\alpha2} } \ ,
\eea

\noindent where $\rho_{\alpha1}$ and $\rho_{\alpha2}$ are
the energy densities of the parts of the fluid $\alpha$
that come from the inflaton and the curvaton decay, respectively,
evaluated in the final state after the end of curvaton decay.
The value $A_{\alpha}=0$ corresponds to fluid $\alpha$ being
produced only in the inflaton decay and the other extreme
$A_{\alpha}=1$ corresponds to $\alpha$ being produced only
in the curvaton decay.

Second, the coefficients $\lambda_{\alpha}$ tell how
efficiently the curvaton decay transfers the original curvaton
perturbation to the decay products. (Similar coefficients exist
for the inflaton decay products, but they are identical for
radiation and matter since the perturbations are adiabatic, and
have been absorbed into $\sR_1$.) If everything 
came from the curvaton, $A_r=A_m=1$, we would have
$\lambda_{\alpha}=\sR_{\alpha}^{out}/\xi_{\sigma}^{in}$.
In other words, multiplying by $\lambda_{\alpha}$ translates
the initial curvaton perturbations $\xi_{\sigma}^{in}$ into the
final radiation and matter perturbations, and the ratio
$\lambda_r/\lambda_m$ measures the relative efficiency of
the transfer.

In section \ref{equations} we noted that when all radiation and
matter come from the curvaton, the decay rates have to obey the limit
$\Gamma_m/\Gamma_r\lesssim 10^{-6}$. In the general case, this
relation is modified by the dilution factors as follows:
\bea
  10^{-6} &\gtrsim& \frac{\rho_m}{\rho_r} \el
  &=& \frac{A_r}{A_m} \frac{\rho_{m2}}{\rho_{r2}} \el
  &\sim& \frac{A_r}{A_m} \frac{\Gamma_m}{\Gamma_r} \ ,
\eea

\noindent where the energy densities are evaluated at the end
of curvaton decay. For $A_m=A_r=1$ we obtain the limit
$\Gamma_m/\Gamma_r\lesssim 10^{-6}$. Decreasing the matter
dilution factor $A_m$ only makes the limit
more stringent, but decreasing $A_r$ relaxes the limit,
so that $\Gamma_m/\Gamma_r$ can be arbitrarily large.

In curvaton models the curvaton spectrum can be non-Gaussian
\cite{Linde:1996, Enqvist:2001, Lyth:2001, Lyth:2002, Sloth:2002, Gordon:2003}.
This is because usually $\sR_1=0$ in \re{Rr}, so that if
$A_r$ is small, $\lambda_r\sR_2$ needs to be large
and the square of the perturbation can become important.
In our model with $\sR_1\neq0$, non-Gaussianity would
appear in the analogous case where we would insist that
the inflaton or curvaton perturbations
make a sizeable contribution to $\sR_r^{out}$
even though $A_r$ is close to 1 or 0, respectively.
(Likewise for $\sR_m^{out}$ and $A_m$.)
Instead, we assume that the perturbations are always small
so that non-Gaussianity is negligible.

The case of \cite{Gupta:2003} is recovered from 
\re{Rr}, \re{Rm}, \re{R} and \re{S} when there is no
matter from the inflaton decay ($A_{m}=1$), the inflaton decay
products have no perturbations ($\sR_1=0$) and matter
inherits the curvaton perturbations with an efficiency of one
($\lambda_m=1$). Then the only remaining parameter is, in the
notation of \cite{Gupta:2003}, $r=A_r\lambda_r$.
The object of the curvaton decay calculation to be presented is
to get the range of the coefficients $\lambda_r$ and $\lambda_m$
in our more general case, and find their dependence on the
parameters of the model.

\subsection{The decay calculation} \label{calculation}

\paragraph{The relevant parameters.}

We will solve the system of background equations
\re{Hubble}, \re{backdens}, \re{backeom}, \re{backsigma} and \re{backtrans}
and perturbation equations \re{pertdens}, \re{perteom}, \re{pertsigma}
and \re{perttrans} numerically,
using the conservation equations \re{backcons} and \re{pertcons}
as a check on the calculation.

Since the calculation is linear in the perturbations, the density
perturbations that radiation and matter have inherited from
the inflaton evolve independently from those that they inherit
from the curvaton. In calculating how the curvaton perturbations
are transferred to radiation and matter, we can therefore put the
inflaton perturbations to zero.

Since the initial velocity of the curvaton field perturbation
is negligible, and the initial field perturbation itself
only sets the amplitude of the perturbations,
the result of the perturbation calculation depends only on the
background dynamics, as in \cite{Gupta:2003}. (This
is quite generally the case, because the perturbations
are governed by linear second order equations which
contain no new parameters apart from the initial
conditions, and the initial velocity is set to zero.)

Matter is initially subdominant, so we can put
the background energy density of the matter originating from the
inflaton (i.e. the initial matter energy density) to zero. The
initial velocity of the curvaton field is negligible, so the result
of the perturbation calculation depends only on the parameters
$m, \Gamma, \Gamma_m/\Gamma_r, \sigma_0$ and $\rho_{r0}$. 
Since the field is stuck until $H\sim m$, the initial energy
density of radiation makes no difference, as long as it dominates.
A larger $\rho_{r0}$ simply means that we start at an earlier
time, and have to wait longer for the curvaton field to become active.

So, the perturbation calculation depends only on the
parameters $m, \Gamma, \Gamma_m/\Gamma_r$ and $\sigma_0$,
specifically on their dimensionless combinations.
In addition to the ratio $\Gamma_m/\Gamma_r$,
we take $\sigma_0/(\sqrt{3} M_{Pl})$ and $\Gamma/m$
as our two independent parameters to vary. In \cite{Gupta:2003},
the dynamics boiled down to a single parameter
in addition to $\Gamma_m/\Gamma_r$
(note that $\Gamma$ was normalised to the initial value of $H$,
so that a given numerical value of $\Gamma$ did not
correspond to a given theoretical model, but depended on the
initial conditions).
The parameter $\sigma_0/(\sqrt{3} M_{Pl})$ controls whether
the curvaton dominates before starting to oscillate (the
case for $\sigma_0/(\sqrt{3} M_{Pl})\gtrsim1$) or not, and
$\Gamma/m$ determines whether the curvaton decays
already as it is starting to oscillate (the case for
$\Gamma/m\gtrsim1$) or later when oscillating.

In the region $\sigma_0/(\sqrt{3} M_{Pl})\ll1, \Gamma/m\ll1$, 
the curvaton oscillates for a long time before dominating or
decaying, and the oscillation is rapid compared to the decay.
The dust approximation should therefore be valid, and we expect
to recover the results of \cite{Gupta:2003}. In contrast, for
$\Gamma/m\gtrsim1$, the curvaton has no time to oscillate
before decaying, so we expect the results to be different
from the dust case (in terms of \re{phisq}, $\gamma_p$ is
not negligible).

\paragraph{Comparison to the dust case.}

From the matter and the curvaton, one can form a composite quantity
which is covariantly conserved for the background \cite{Gupta:2003}:
\bea
  \rho_{comp} = \rho_m + \frac{\Gamma_m}{\Gamma} \rho_{\sigma} \el
  P_{comp} = \frac{\Gamma_m}{\Gamma} P_{\sigma} \ .
\eea

When the curvaton behaves like dust, this quantity has a (trivially)
barotropic equation of state, $P_{comp}=P_{comp}(\rho_{comp})=0$,
so the corresponding curvature perturbation
is conserved. To study the goodness of the dust
approximation, we therefore introduce the variable
\bea \label{xicomp}
  \xi_{comp} \equiv - \frac{\delta\rho_{comp}}{3\rho_{comp}} \ ,
\eea

\noindent using in analogy with \re{xisigma} the letter
$\xi$ to refer to a quantity which is a conserved curvature
perturbation only when the curvaton behaves like dust. Before the
curvaton decays, we have $\rho_m=0$ (recall that in this
calculation we have no matter from the inflaton), so that
$\xi_{comp}=\xi_{\sigma}$, and after the
decay is over we have $\rho_{\sigma}=0$, so that
$\xi_{comp}=\sR_m$. So, $\xi_{comp}$ is conserved
both before and after the decay. The ratio of the final
and initial values is
$\xi_{comp}^{out}/\xi_{comp}^{in}=\sR_m^{out}/\xi_{\sigma}^{in}=\lambda_m$.
As noted, when the curvaton behaves like dust, $\xi_{\sigma}$
is conserved throughout and we have $\lambda_m=1$ as found in 
\cite{Gupta:2003}. The value of $\lambda_m$ measures the
conservation of $\xi_{comp}$ and thus the goodness of the dust
approximation.

\paragraph{The numerical calculation.}
 
We vary $\sigma_0/(\sqrt{3} M_{Pl})$ in the range $[10^{-4},10]$,
going from the curvaton being completely subdominant
to completely dominating when it is starting to oscillate
(but avoid a long period of curvaton-driven inflation).
We vary $\Gamma/m$ in the range $[10^{-6},10]$, going from the
curvaton decaying deep in the oscillating regime to decaying
just as it is starting to oscillate.
We take 100 logarithmically spaced steps for each parameter,
covering 10 000 models in all.
We start our integration when $H/m\gg1$, the
universe is dominated by radiation and the curvaton is frozen at the
initial field value $\sigma_0+\delta\sigma_0$. For a given value of
the parameters we evolve the system until the curvaton has decayed,
and we are left with radiation and matter only.

We have studied a number of the above models with different values
of the ratio $\Gamma_m/\Gamma_r$ in the range $[10^{-6},10^6]$. The
results depend only weakly on the ratio,
so we have chosen not to do a systematic scan like for the other
two parameters. The behaviour shown in the pictures below would
be qualitatively the same for any value of $\Gamma_m/\Gamma_r$
in the above range.

There are two reasons why the
ratio $\Gamma_m/\Gamma_r$ doesn't make much difference to the perturbations.
First, the behaviour of the background is not affected by how much
matter the decay produces since we assume that matter is completely
subdominant throughout. Second, changing $\Gamma_m/\Gamma_r$
changes the relative production of radiation and matter for
both the background energy densities $\rho_{\alpha}$ and the
perturbations $\delta\rho_{\alpha}$, so the perturbation
variables which are given by their ratios
(for example, $\sR_m^{out}=-\delta\rho_m/(3\rho_m)$)
are only weakly affected.

\begin{figure*}[t]
\includegraphics[clip=true,width=0.9\textwidth]{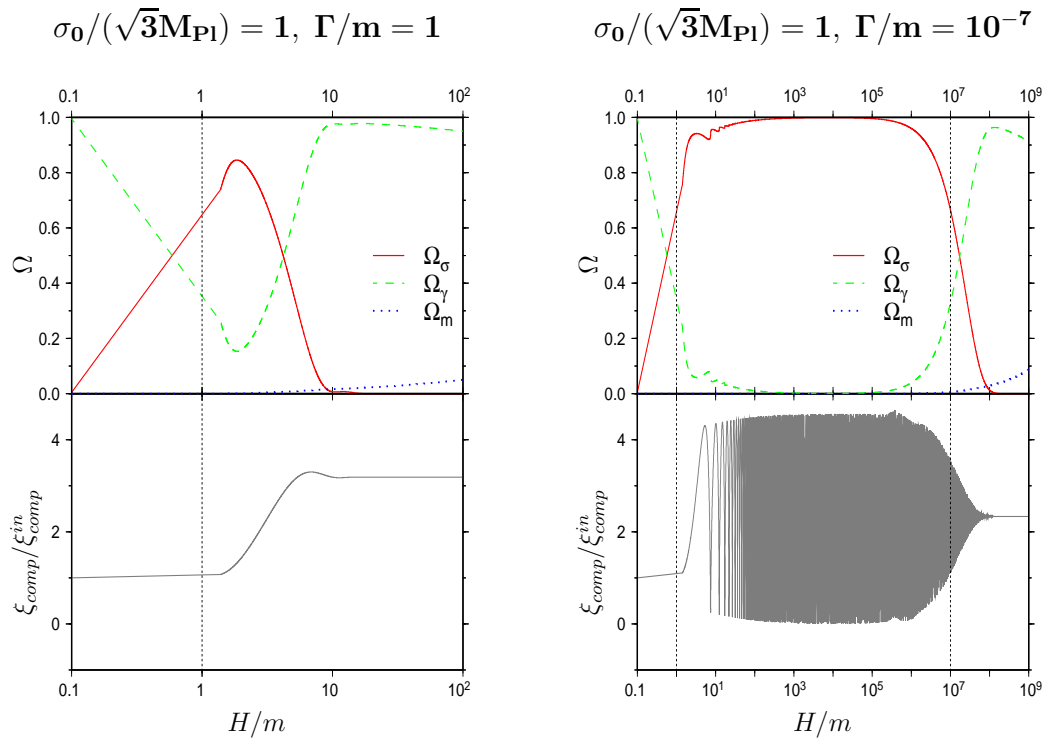}
\caption{Background energy densities (up) and $\xi_{comp}$ (down) for 
a large initial value of the field. On the left, a model
where the field decays just as it is starting to oscillate
and on the right, a model where the field oscillates before decaying. The
dotted vertical lines mark the nominal start of oscillation and decay.}
\label{figure:largefield}
\end{figure*}

\begin{figure*}[t]
\includegraphics[clip=true,width=0.9\textwidth]{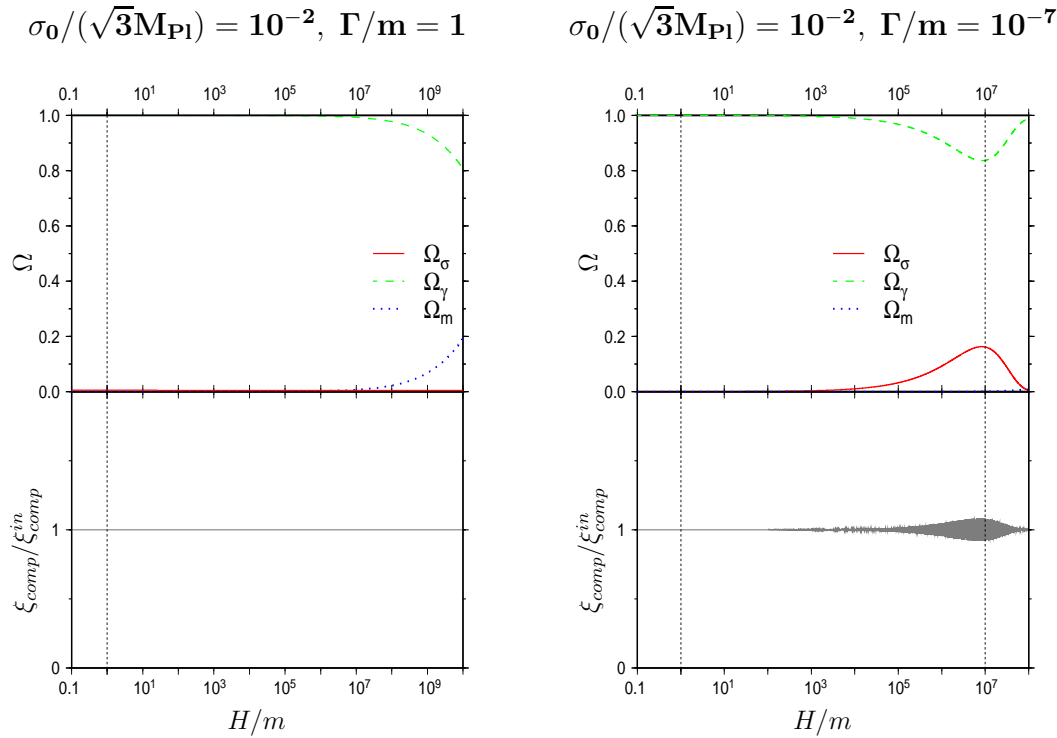}
\caption{Background  energy densities (up) and $\xi_{comp}$ (down) for a 
small initial value of the field.}
\label{figure:smallfield}
\end{figure*}

\begin{figure*}[tb!]
\includegraphics[clip=true,width=0.8\textwidth]{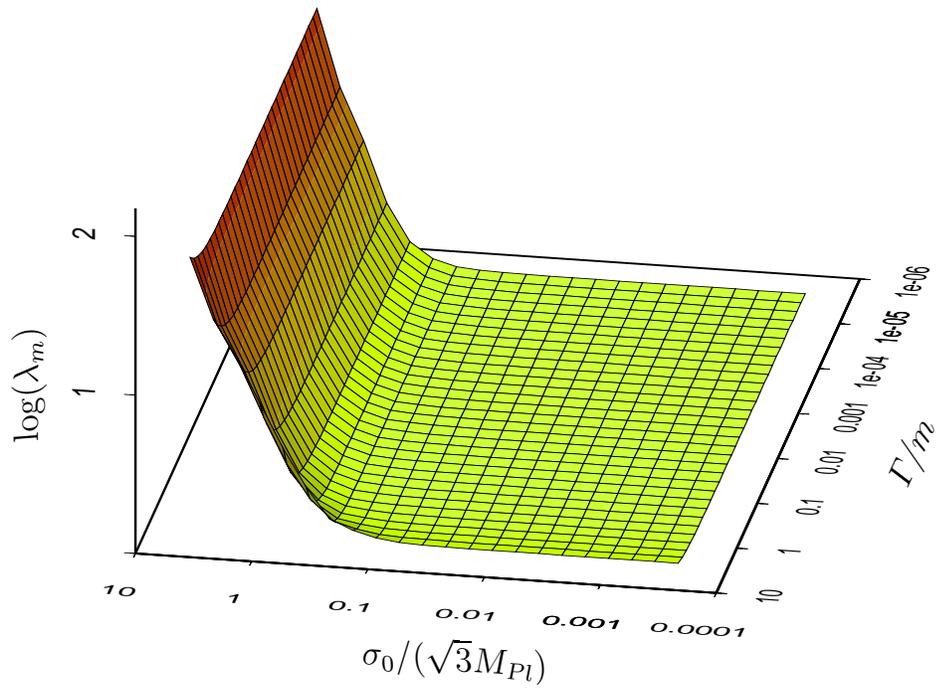}
\caption{The matter decay coefficient $\lambda_m$ on a logarithmic scale.}
\label{figure:lambdam}
\end{figure*}

\begin{figure*}[tb!]
\includegraphics[clip=true,width=0.8\textwidth]{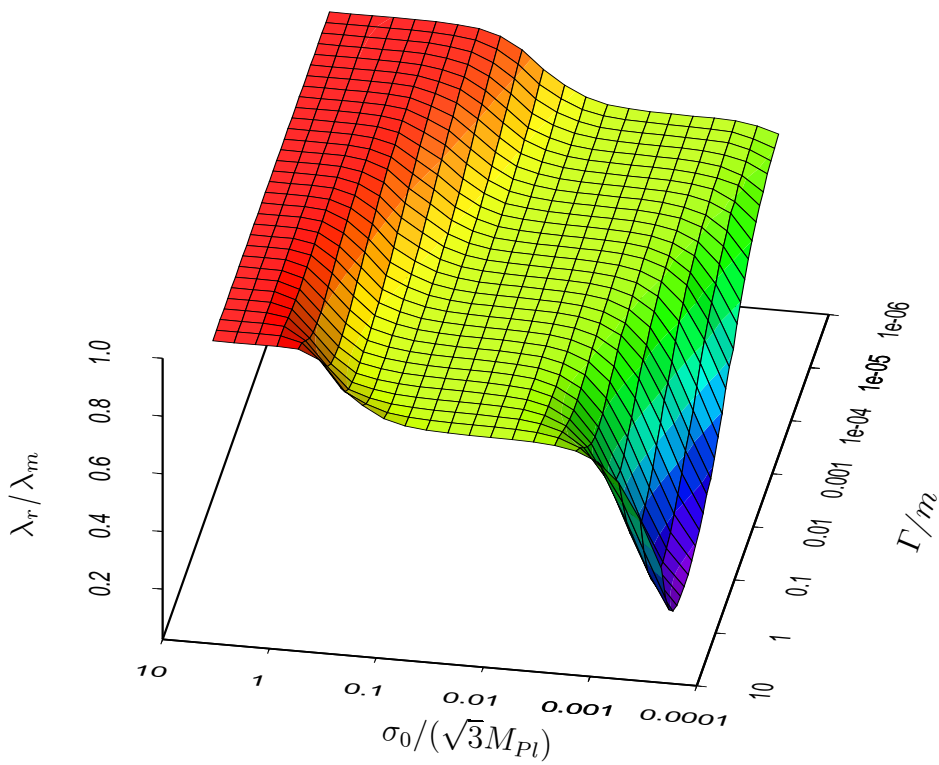}
\caption{The ratio of the decay coefficients for radiation and matter,
$\lambda_r/\lambda_m$.}
\label{figure:lambdarm}
\end{figure*}

The results of the calculation are shown in Figures
\ref{figure:largefield} to \ref{figure:lambdarm}. In Figures
\ref{figure:largefield} and \ref{figure:smallfield}
we show the evolution of the background energy densities
and the perturbation variable $\xi_{comp}$.
In Figure \ref{figure:largefield}, we have chosen a
large initial field value, so
the curvaton contributes significantly to the energy density
when it decays. On the left, the curvaton has a large decay rate,
so that it decays before really starting to oscillate. On the
right, the decay rate is small, so that the curvaton oscillates
rapidly at the time of decay and thus resembles a pressureless
fluid more closely. In Figure \ref{figure:smallfield}, we
show the analogous plots for a small initial field value, when the curvaton
does not contribute significantly to the background before
starting to oscillate. Note that even in the model on the
left where the curvaton decays before properly oscillating,
we recover the dust result $\lambda_m=1$.

The surprising result that the dust approximation
is valid even when the field does not oscillate when it
decays is a general feature, as seen from Figure \ref{figure:lambdam},
which shows $\lambda_m$ for the whole of our parameter space.
The value of $\lambda_m$ is practically independent of the decay
rate $\Gamma/m$: as long as the initial field value
$\sigma_0/(\sqrt{3} M_{Pl})$ is less than 0.1 (so that
the curvaton does not dominate before oscillating), 
$\lambda_m$ is very close to 1, even when the field decays
without any oscillations. On the other hand, for larger values
of $\sigma_0/(\sqrt{3} M_{Pl})$, $\lambda_m$ rises rapidly, up to a
maximum of about 350 at the edge of our parameter space.

For fitting our spectrum \re{R}, \re{S} to the data, we need
to know, in addition to $\lambda_m$, the ratio
$\lambda_r/\lambda_m$ which measures the relative efficiency
of the transfer of the curvaton perturbations to radiation
and matter. In Figure \ref{figure:lambdarm} we show 
$\lambda_r/\lambda_m$ for our parameter space.
Like $\lambda_m$, the ratio $\lambda_r/\lambda_m$ is
more sensitive to the initial field value
$\sigma_0/(\sqrt{3} M_{Pl})$ than to the decay rate $\Gamma/m$,
though it does grow with decreasing $\Gamma/m$.
For small values of the field, the ratio approaches 0
rapidly, and for large values it tends to 1. The
upper limit is easy to understand: for large $\sigma_0$,
the curvaton completely dominates when it decays, so its
perturbations are adiabatic, and the perturbations of its
decay products have to be adiabatic as well \cite{Weinberg:2004a},
so radiation and matter inherit the same perturbations.

We have studied the validity of the dust approximation, and have
surprisingly found that its results for the decay are valid
even when the field does not oscillate when it decays.
We have also found the range of the parameters $\lambda_m$ and
$\lambda_r/\lambda_m$, which we need for fitting our spectrum
to CMB data.

\section{The fit to WMAP data} \label{WMAP}

\subsection{The spectrum of perturbations} \label{spectrum}

Having found the ranges for the parameters for the mixture
of inflaton and curvaton perturbations in the spectrum
\re{R}, \re{S}, let us now fit the spectrum to the WMAP data
and see what such a mixture implies for the CMB anisotropies.

We have not yet specified the inflaton and curvaton spectra $\hsR_1$
and $\hsR_2$ (in this section we introduce a hat to denote
random variable). As the perturbations are assumed to be generated
during inflation, the spectra depend on the inflationary model.
We will not choose a particular model, but assume that
the spectra are given by simple power laws,
\bea
  {\textstyle\left(\frac{k^3}{2\pi^2}\right)^{\frac{1}{2}}} \hsR_1 &=& N_1 \kt^{\frac{n_1-1}{2}} \ahat_1(\bk) \el
  {\textstyle\left(\frac{k^3}{2\pi^2}\right)^{\frac{1}{2}}} \hsR_2 &=& N_2 \kt^{\frac{n_2-1}{2}} \ahat_2(\bk) \ ,
\eea
where $N_1$ and $N_2$ are constant amplitudes, $n_1$ and $n_2$ are
constant spectral indices and $\kt=k/k_0$ with
$k_0$ = 0.05 Mpc$^{-1}$ is the dimensionless wavenumber.
We have denoted by $\ahat_1(\bk)$ and $\ahat_2(\bk)$
Gaussian random variables which obey
\bea
  \langle \ahat_1 \rangle = \langle \ahat_2 \rangle =0 \ ,
\quad \langle \ahat_i(\bk)\ahat_j^*(\bk') \rangle = 
\delta_{ij}\delta^{(3)}(\bk-\bk') \ .
\eea

If the perturbation spectra were generated during a period of
slow-roll inflation, the spectral indices would be (to first
order in the slow-roll parameters) \cite{Lyth:2001, Liddle:2000}
\bea \label{slowroll}
  n_1 = 1 + 2\eta_1 - 6\epsilon_H \el
  n_2 = 1 + 2\eta_2 - 2\epsilon_H \ ,
\eea

\noindent where $\epsilon_H=-\Hdot/H^2$, and $\eta_1$ and
$\eta_2$ are the slow-roll parameters associated with the
curvature of the potential in the inflaton and curvaton
directions, respectively,
\bea
  \eta_1 &=& M_{Pl}^2 \frac{1}{V}\frac{\pat^2 V}{\pat\varphi^2} \el
  \eta_2 &=& M_{Pl}^2 \frac{1}{V}\frac{\pat^2 V}{\pat\sigma^2} =\frac{m^2}{3 H^2} \ ,
\eea

\noindent where $\varphi$ is the inflaton field and
$V$ is the inflaton-curvaton-potential
$V(\sigma,\varphi)=\frac{1}{2}m^2\sigma^2 + V(\varphi)$
(we neglect inflaton-curvaton interactions).
Since the mass of the curvaton is assumed to be much
smaller than the Hubble parameter during inflation,
$\eta_2\ll1$, and since during slow-roll typically
$|\Hdot|\ll H^2$, one might expect $n_2$ to be only marginally
smaller than 1, though $n_1$ could be more different from
(and possibly larger than) 1 due to a non-negligible
$\eta_1$.

However, in the curvaton scenario there is no need to assume
that inflation is slow-roll \cite{Dimopoulos:2002} or even
that inflationary dynamics are driven by a scalar field
\cite{Starobinsky:1980, Brandenberger:2003, Brandenberger:2004}.
Also, even in single-field inflation the spectral index can
be very different from one, while having little running
\cite{Garcia-Bellido:1996, Kinney:1997, Kinney:2003}.
We do not limit ourselves to slow-roll inflation, and
keep the spectral indices as free parameters with a wide range
(and do not introduce running for the indices).

In section \ref{decay}, we calculated the decay of the curvaton
into two components, radiation and matter, which inherited
the curvaton spectrum with different efficiency coefficients
$\lambda_{\alpha}$. In the fit
to the CMB, we have four components: baryons, CDM,
(massive) neutrinos and photons.

The CDM is assumed to be non-relativistic during the decay,
and will therefore inherit the curvaton spectrum with the
efficiency $\lambda_m$, so its spectrum of
perturbations is given by \re{Rm}. The photons will
inherit the curvaton perturbations with the efficiency $\lambda_r$,
and have the spectrum \re{Rr}. During curvaton decay,
neutrinos are also relativistic, but since the neutrino
dilution factor $A_{\nu}$ can be different from the
photon dilution factor $A_{\gamma}$, neutrinos and photons
could get a different spectrum of perturbations. However, we
assume that there is no net lepton number, in which case the
neutrinos and photons end up having the same spectrum
by virtue of being in thermal equilibrium
\cite{Lyth:2002, Bucher:1999, Weinberg:2004b}, so there are
no neutrino isocurvature perturbations.

Whether the baryons inherit the curvaton spectrum with the
efficiency factor for radiation or matter depends
on whether the degrees of freedom carrying baryon number are
relativistic or non-relativistic during curvaton decay. In any case,
there is no reason for the baryon dilution factor $A_b$ to be
the same as the CDM or photon dilution factors, so there will
in general be baryon-CDM and baryon-photon isocurvature perturbations.
However, since baryon and CDM isocurvature perturbations behave
similarly \cite{Gordon:2002, Bucher:1999}, the isocurvature
perturbation between baryons and CDM can be transformed to zero,
at the price of adding more isocurvature between CDM and photons.
Then the only component with isocurvature perturbations is the
CDM, and the spectrum reads
\bea \label{spectrum1}
  \fl {\textstyle\left(\frac{k^3}{2\pi^2}\right)^{\frac{1}{2}}} \hsR^{out}_{\gamma} &=& (1-A_{\gamma}) N_1 \kt^{\frac{n_1-1}{2}} \ahat_1(\bk) +
A_{\gamma}\lambda_r N_2 \kt^{\frac{n_2-1}{2}}  \ahat_2(\bk) \el
  \fl {\textstyle\left(\frac{k^3}{2\pi^2}\right)^{\frac{1}{2}}} \hsS_{cdm,\gamma}^{out} &=& 3 [ (A_{cdm}-A_{\gamma}) + x_b (A_b-A_{\gamma}) ] N_1 \kt^{\frac{n_1-1}{2}} \ahat_1(\bk) \el
  && - 3 [ (A_{cdm}\lambda_m-A_{\gamma}\lambda_r) + x_b (A_b\lambda_b-A_{\gamma}\lambda_r) ] N_2 \kt^{\frac{n_2-1}{2}} \ahat_2(\bk) \ ,
\eea

\noindent where $x_b\equiv\rho_b/\rho_{cdm}$
and $\lambda_b$ is the baryon decay coefficient,
which can be either $\lambda_r$ or $\lambda_m$.

The most general form of a spectrum with an adiabatic
plus one isocurvature mode, both of which are a
combination of two power laws, has six independent
parameters: two spectral indices and four amplitudes (one for
each of the two different powers of $\kt$ for the adiabatic and the
isocurvature mode). The expression \re{spectrum1} is therefore
overdetermined, and though the theoretical interpretation of
the parameters is clear, their relation to observable features
is murky. We will write the spectrum \re{spectrum1} in a form more
suited for comparison with observations:
\bea \label{spectrum2}
 {\textstyle\left(\frac{k^3}{2\pi^2}\right)^{\frac{1}{2}}} \hsR^{out}_{\gamma} &=& N_{\sR} \left( \cos\phi\, \kt^{\frac{n_1-1}{2}} \ahat_1(\bk)  + \sin\phi\, \kt^{\frac{n_2-1}{2}}\ahat_2(\bk)  \right) \el
 {\textstyle\left(\frac{k^3}{2\pi^2}\right)^{\frac{1}{2}}} \hsS_{cdm,\gamma}^{out} &=& N_{\sR} f_{\sS} \left( \cos\theta\, \kt^{\frac{n_1-1}{2}} \ahat_1(\bk)  - \sin\theta\, \kt^{\frac{n_2-1}{2}} \ahat_2(\bk) \right) \ ,
\eea

\noindent where the new constant parameters
$N_{\sR}, f_{\sS}, \phi, \theta$ can be written in terms of
the old parameters in \re{spectrum1} (the inverse is obviously
not true, since \re{spectrum1} has more parameters than
\re{spectrum2}).

The new parameters have a transparent interpretation in terms of
observable quantities: $N_{\sR}$ is the amplitude of the adiabatic
perturbations, the isocurvature fraction $f_{\sS}$
measures the amplitude of the isocurvature
perturbations relative to the adiabatic ones, and the angles $\phi$ and
$\theta$ determine the proportion of the adiabatic and isocurvature
perturbations, respectively, that have the spectral index $n_1$
or $n_2$. 

The generality of the spectrum \re{spectrum2} goes beyond the
inflaton-curvaton model that we have discussed. It is the most general
form for the case when baryons, CDM and photons inherit power-law
perturbations from two independent sources. For example, it
could describe a double inflation model\footnote{Note that even
in the slow-roll case \re{slowroll} the spectrum \re{spectrum2}
does not satisfy the consistency relations presented in
\cite{Wands:2002} for two-field inflation, except when
$\cos\theta=0$. In that case, either the inflaton perturbations
are zero, or all fluids come from the inflaton and curvaton
with the same ratio $A_{\alpha}$ (barring a
fortuitous cancellation between baryons and CDM in \re{spectrum1}).}.
Also, in realistic supersymmetric GUT theories, cosmic strings are
typically produced after inflation, leading to isocurvature
perturbations \cite{Jeannerot:2003}.
The isocurvature perturbations produced by cosmic strings are
usually considered to be uncorrelated with the pre-existing
adiabatic perturbations from the inflationary era, but if they
were correlated and the strings would decay, a spectrum like
\re{spectrum2} might result.

In the general case, the ranges for the parameters in \re{spectrum2}
are as follows. The amplitude $N_{\sR}$ is positive, 
the isocurvature fraction $f_{\sS}$ can be negative,
positive or zero, the angle $\phi$ is from the interval $[0, \pi/2]$
and the angle $\theta$ is from the interval $[0, \pi]$.

Note that the roles of the inflaton and the curvaton are
completely symmetric: the spectrum
\re{spectrum2} is invariant under the transformation
$n_1 \leftrightarrow n_2$, $\cos\phi \leftrightarrow \sin\phi$,
$\cos\theta \leftrightarrow \mbox{sign}(\cos\theta)\sin\theta$
and $f_S \rightarrow -\mbox{sign}(\cos\theta)f_S$. To avoid
scanning the parameter space twice,
we break the symmetry by writing $n_1 = n_2 + \Delta n$ and
studying only models where the inflaton spectral index is
larger than or equal to the curvaton spectral index, $\Delta n\ge0$.

If the transfer of curvaton perturbations to matter and
radiation were equally efficient, $\lambda_r=\lambda_m$,
then $\sin\theta$ and $\cos\theta$ would necessarily have the
same sign, so that $\theta$ would be restricted to the range
$[0, \pi/2]$. In our model the transfer to matter is more efficient
than the transfer to radiation, $\lambda_r<\lambda_m$,
so that $\theta$ has the full range $[0, \pi]$ when $f_\sS>0$,
but is restricted to the range $[0, \pi/2]$ when $f_\sS<0$.
So, the role of the decay coefficients $\lambda_{\alpha}$
is to expand the range of $\theta$ in the spectrum \re{spectrum2}.
In our analysis we will allow $\theta$ to vary between $[0, \pi]$,
and at the end remove the forbidden region $f_\sS<0, \theta>\pi/2$.

\subsection{Angular power and (anti)correlation} \label{anticorrelation}

\paragraph{The angular power spectra.}

The temperature-temperature (TT) angular power spectrum resulting from the
primordial spectra \re{spectrum2} is
\bea \label{Cl}
  C_l^{TT} &=& 4\pi \int\frac{\rmd k}{k} \frac{k^3}{2\pi^2} \langle |g_{l,\sR}^T(k)\hsR(\bk)+g_{l,\sS}^T(k)\hsS(\bk)|^2\rangle \el
  &=& 4\pi \int\frac{\rmd k}{k}  \frac{k^3}{2\pi^2} \left(g_{l,\sR}^T(k)^2\langle |\hsR(\bk)|^2\rangle + g_{l,\sS}^T(k)^2\langle |\hsS(\bk)|^2\rangle \right. \el
  && \left. + g_{l,\sR}^T(k) g_{l,\sS}^T(k) \langle \hsR(\bk)\hsS(\bk)^* + \hsR(\bk)^*\hsS(\bk)\rangle \right) \el
  \label{Cla} &=& 4\pi \int\frac{\rmd k}{k}  N_{\sR}^2 \left[ g_{l,\sR}^T(k)^2\left( \cos^2\phi\,\kt^{n_1-1} + \sin^2\phi\,\kt^{n_2-1} \right) \right. \el
  && + f_{\sS}^2\, g_{l,\sS}^T(k)^2\left( \cos^2\theta\,\kt^{n_1-1} + \sin^2\theta\,\kt^{n_2-1} \right) \el
  && \left. + 2 g_{l,\sR}^T(k) f_{\sS} g_{l,\sS}^T(k) \left( \cos\phi\cos\theta\,\kt^{n_1-1} - \sin\phi\sin\theta\,\kt^{n_2-1} \right) \right] \\
  \label{Clb} &=& 4\pi \int\frac{\rmd k}{k}  N_{\sR}^2 \left[\left(g_{l,\sR}^T(k)\cos\phi + f_{\sS} g_{l,\sS}^T(k) \cos\theta\right)^2 \kt^{n_1-1} \right. \el
  && \left. + \left(g_{l,\sR}^T(k)\sin\phi - f_{\sS} g_{l,\sS}^T(k) \sin\theta\right)^2 \kt^{n_2-1} \right] \ ,
\eea

\noindent where $g_{l,\sR}^T(k)$ and $g_{l,\sS}^T(k)$ are the
functions which transfer the primordial curvature and entropy
perturbations of the mode with the wavenumber $k$
from the early radiation-dominated era to the present temperature
fluctuation at the multipole $l$.
Written like this, the role of $f_{\sS}$ in scaling 
the amplitude of the isocurvature modes via $g_{l,\sS}^T(k)$ is
particularly transparent.

For the temperature-polarisation (TE) cross-correlation power
spectrum we have
\bea \label{ClTE}
  \fl C_l^{TE} = 4\pi \int\frac{\rmd k}{k} \frac{k^3}{2\pi^2} \langle [g_{l,\sR}^T(k)\hsR(\bk)+g_{l,\sS}^T(k)\hsS(\bk)]^* \times
 [g_{l,\sR}^E(k)\hsR(\bk)+g_{l,\sS}^E(k)\hsS(\bk)]
\rangle \ ,
\eea

\noindent where  $g_{l,\sR}^E(k)$ and $g_{l,\sS}^E(k)$ are the transfer
functions for the polarisation $E$-mode (for
details see e.g. \cite{Valiviita:2003a}).

There are six different terms in \re{Cla}. The first two
are due to the primordial adiabatic perturbations $\hsR^{out}_{\gamma}$,
and they will be denoted by $C_l^{\rm adi1}$ and $C_l^{\rm adi2}$.
The next two terms are due to the primordial isocurvature
perturbations $\hsS^{out}_{cdm,\gamma}$, and they will
be denoted by $C_l^{\rm iso1}$ and $C_l^{\rm iso2}$.
Finally, the last two terms arise from the
correlation between adiabatic and isocurvature
perturbations, and will be denoted by $C_l^{\rm cor1}$ and
$C_l^{\rm cor2}$. The same notation will be used
for the components of the TE power spectrum.

\paragraph{The role of CDM isocurvature perturbations.}

The acoustic peak structure in the measured
angular power spectrum looks distinctly adiabatic.
CDM isocurvature perturbations have valleys where
adiabatic perturbations have peaks and are also damped
faster with increasing $l$. Therefore pure CDM isocurvature
perturbations cannot fit the data and are ruled out
\cite{Enqvist:2001fu}.
However, a mixture where adiabatic perturbations
dominate the peak structure and CDM isocurvature
perturbations give a significant contribution at
low multipoles remains an interesting possibility.

At low multipoles the Sachs-Wolfe effect gives
$g_{l,\sS}^T(k) = -2 (1-f_{\nu}) g_{l,\sR}^T(k)$\footnote{The
factor $1-f_{\nu}$ arises because we are considering the
cold dark matter-photon isocurvature perturbation
$\hsS_{cdm, \gamma}$ instead of the total dark
matter-photon isocurvature perturbation $\hsS_{dm, \gamma}$.
This issue will be discussed in section \ref{likelihood}.}.
The rapid damping of $g_{l,\sS}^T(k)$ with increasing
$l$ means that if the primordial adiabatic and
isocurvature perturbation amplitudes are
of the same order,
the isocurvature perturbations modify only the low-$l$ part
(practically the Sachs-Wolfe plateau) of the
final $C_l$ spectrum observed today.
This is natural in our model where baryons, CDM and photons
are all generated by both inflaton and curvaton decay.
In terms of the spectrum \re{spectrum2}, this means
$f_\sS\sim1$, which is the generic value. To get a
much larger or smaller value, one
has to tune the model in some way (say, by having
$A_{\gamma}=A_{cdm}=A_{b}$ to a high accuracy).

Uncorrelated adiabatic and isocurvature perturbations
add power on the SW plateau relative to the acoustic
peaks, and since the observations indicate a deficit
of power, this does not improve the fit to the data \cite{Enqvist:2000hp}.
However, in our model the adiabatic and isocurvature perturbations
are correlated. Negative correlation%
\footnote{Note that in our convention negative correlation for
the $C_l$:s on the SW plateau corresponds to positive correlation
for $\hsR$ and $\hsS$, since
$g_{l,\sS}^T(k) = -2 (1-f_{\nu}) g_{l,\sR}^T(k)$. One should be
careful, as sign conventions vary. Some authors prefer to use
$\hat\zeta=-\hsR$, and others define the variables with
a different sign; e.g. in \cite{Valiviita:2003a} there was
an additional minus sign in the definition of $\hsR$, so that
negative correlation between $\hsR$ and $\hsS$ corresponded
to a negative $C_l^{\rm cor}$.}
will naturally suppress the power on the SW plateau
and thus improve the fit. The suppression is large if
$\hsR\sim\hsS$, and total when $\hsR=2(1-f_{\nu})\hsS$. Again,
note that our model naturally yields $\hsR\sim\hsS$ which
is required for significant cancellation.
Also, as  we will discuss in section \ref{suppression},
our spectrum \re{spectrum2} gives more freedom
to adjust the cancellation of the low multipoles than in previous
studies of suppression due to anticorrelation
\cite{Moroi:2001, Moroi:2002, Langlois:1999a, Valiviita:2003a, Langlois:1999b, Amendola:2001, Peiris:2003, Moroi:2003, Valiviita:2003b, Gordon:2004ez}.

We note in passing that the conclusion that the peak structure
excludes a sizeable contribution from CDM isocurvature modes
outside of the SW plateau does not hold
for neutrino density and neutrino velocity isocurvature
modes, since their peak structure is similar
to that of adiabatic modes \cite{Bucher:1999, Bucher:2000kb}.
In fact, there is an explicit example where
isocurvature modes give a major contribution
to all of the observed acoustic peaks \cite{Bucher:2004}.

\subsection{Fitting method}

We fit the spectra \re{Cla} and \re{ClTE} to the WMAP first-year
data \cite{Kogut:2003et, Hinshaw:2003ex, Verde:2003ey}
by running several Monte Carlo chains on a supercomputer.
We vary the following 11 {\it primary parameters} (parameter
ranges are given in Table \ref{table:parameters}).
We have five standard parameters related to the cosmological
background, called the {\it hard parameters:}
the physical baryon density $\omega_b=h^2\Omega_b$,
the physical dark matter density $\omega_{dm}=h^2\Omega_{dm}$,
the ratio of the sound horizon to the angular diameter distance
(i.e. the acoustic peak scale in radians) $\Theta$,
the optical depth due to reionisation $\tau$ and
the neutrino fraction $f_\nu=\omega_\nu/\omega_{dm}$.
We also assume that there is vacuum energy, the density of which
is fixed by the constraint that the universe is spatially flat,
$\Omega_\Lambda + \Omega_m = 1$, where
$\Omega_m = \Omega_b + \Omega_{dm}$.
We have six parameters for the spectrum of
perturbations, instead of the usual two: the spectral index
$n_1$, the difference between the spectral indices
$\Delta n=n_1-n_2$, the overall normalisation
$\ln(10^{10} N_\sR^2)$, the fraction of adiabatic
perturbations that come from the inflaton $\cos\phi$,
the fraction of isocurvature perturbations that come from the
inflaton $\cos\theta$, and the
CDM-photon isocurvature fraction $f_\sS$.

The technical details of our analysis are very similar to
those described in the appendix of \cite{Tegmark:2003ud}.
We use a modified version of CosmoMC \cite{cosmomc} to
generate 40 Monte Carlo Markov Chains that start from
randomly chosen points in the parameter space. The transfer
functions $g_{l,\sR}^T(k)$,  $g_{l,\sS}^T(k)$,  $g_{l,\sR}^E(k)$,
and  $g_{l,\sS}^E(k)$ are calculated with our modified version
of CAMB \cite{camb} whenever any hard parameter is changed. The
integrals \re{Cla} and \re{ClTE} are then evaluated for each
model. After running between 8 and 12 days, each chain contains
about 10 000 accepted steps (150 000 step trials). We cut off
burn-in periods and neglect 13
chains that never burned-in but instead got stuck
in a local likelihood minimum. The correlation length
(i.e. the number of steps between independent samples) is
quite long in our chains, typically 170 -- 1000. Nevertheless,
after the described procedure, we still have over
220 000 accepted steps (3 300 000 step trials) giving  
7500 independent samples to analyse.

In the same way, we create chains for a $\Lambda$CDM
model with adiabatic power-law perturbations
(and without massive neutrinos) to serve as a reference point.
The 6 primary parameters of the reference model are $\omega_b$,
$\omega_{dm}$, $\Theta$ and $\tau$ for the background and
$n$, $\ln(10^{10} N_\sR^2)$ for the perturbations.

We produce a 1-dimensional marginalised likelihood distribution
(probability density)
for each primary parameter and for some {\it derived parameters}
such as
the spectral index $n_2$,
the dark matter-photon isocurvature fraction $\tilde f_{\sS}$,
the vacuum energy density $\Omega_\Lambda$,
the baryon energy density $\Omega_b$,
the matter energy density $\Omega_m$,
the CDM energy density $\Omega_{cdm}$,
the (massive) neutrino energy density $\Omega_\nu$,
the age of the universe,
the reionisation redshift $z_{re}$ and
the Hubble parameter today $H_0=h\times 100$ km/s/Mpc.

We do not apply any priors to primary parameters,
but use a top-hat prior $0.21<h<1.00$ for the
Hubble parameter and $\Omega_\Lambda\ge0$ for the
vacuum energy density.
The main focus of the present paper is the study of the
perturbation spectrum in the mixed inflaton-curvaton model,
so we will here simply do a fit to the WMAP data,
and briefly discuss the interesting features of the fit.
We will follow up with a more detailed analysis of cosmological
parameter evaluation in a separate paper \cite{Ferrer:2004},
where we will also include data from high-$l$ CMB
measurements and the Sloan Digital Sky Survey (SDSS)
\cite{Tegmark:2003ud, Tegmark:2003uf}.

\subsection{Results of the data fit} \label{fitresults}

There are 1348 WMAP data points and our model has 11 parameters,
so the reduced number of degrees of freedom is $\nu = 1348 - 11 = 1337$.
The reference pure adiabatic $\Lambda$CDM model has 6
parameters giving $\nu = 1342$. Our best-fit model has
$\chi^2 = 1423.9$ and the best-fit reference model
has $\chi^2 = 1429.0$. In the first two columns of 
Table \ref{table:parameters} we compare the parameters
of these models. Even though our model has a better $\chi^2$
value, $\chi^2/\nu$ turns out to be exactly the same
(1.065) for both models. Thus this measure-of-goodness does
not support introducing the isocurvature degrees of freedom\footnote{As
discussed in \cite{Liddle:2004}, the criteria
for introducing new parameters should in fact be more strict
than simply getting a slightly better $\chi^2/\nu$.}.
However, we are interested in the SW plateau (in particular
the quadrupole power) beyond its statistical weight
in the overall fit. On the last three lines of
Table \ref{table:parameters} we give the quadrupole
power of both models as well as the $\chi^2$ of the first
two data points (quadrupole and octopole) and their
contribution to the total $\chi^2$. The quadrupole
power in our best-fit model is 125 $\mu$K$^2$ smaller than
in the $\Lambda$CDM model. This makes our fit better in
the SW plateau and drops the $\chi^2$ contribution of the
first two data points from 0.51\% to 0.45\%.

The parameters of the best-fit model are
somewhat exotic. The physical baryon density
$\omega_b=0.041$ is much larger than the value
given by BBN \cite{Sarkar:2002er}. The
large neutrino density $f_{\nu}=0.86$ implies
(along with $\omega_{dm}=0.12$) that the sum of
neutrino masses is 9.2 eV, in conflict with
measurements of tritium decay which give an upper limit
of $m_{\nu_e}=2.2$ eV for the electron neutrino mass
\cite{Kraus:2003}, yielding an upper limit of 6.6 eV for
the sum of neutrino masses (when combined with the
small mass differences from oscillation experiments).
Note also the very small CDM density,
$\Omega_{cdm}=0.028$. One would expect this to be highly
problematic for structure formation, but on the other
hand the spectral index on the relevant scales is
deeply blue, $n_1=3.9$, which significantly enhances
the amplitude of perturbations on the scales probed by 
the measurements of large-scale structure.

The best-fit isocurvature model (for which we do not give
a $C_l$ plot) is not meant to be taken as an alternative
to the best-fit adiabatic power-law $\Lambda$CDM model.
Rather, it underlines the point that models with radically
different background cosmologies and perturbation spectra can
fit the WMAP data even slightly better than the standard model.
Including CDM isocurvature perturbations opens up the space of
well-fitting models, and the cosmological parameters in these models
are no longer clustered around the best-fitting model.
The addition of priors from other cosmological observations,
as well as data on the power spectrum from
high-$l$ CMB measurements and large-scale structure
will exclude many of these models. For example, our best-fit
model does not fit the data from high-$l$ CMB measurements:
the nearly scale-free spectral index $n_2 = 0.988$
dominates until the third acoustic peak, but the
large spectral index $n_1 = 3.906$ gives too
much power in the region $l>900$.

For now, the message to take away is that one cannot
determine the cosmological parameters from the WMAP data
alone with any degree of confidence without making strong
assumptions about the primordial power spectrum. Turning this
around, it is not possible to determine the primordial power
spectrum using only the WMAP data without making strong assumptions
about the cosmological parameters. This has been previously emphasised
in \cite{Trotta:2001, Trotta:2002}.
As a warning example, we note that
the original implementation of the pre-big bang scenario
\cite{Brustein:1994kn} was ruled out in part because the spectral
index $n=4$ of the adiabatic perturbations coming from the
dilaton was too large to fit the CMB data, and the second
version \cite{PBB2} was ruled out because the axion with a
spectral index $n=1$ carried only isocurvature perturbations.
Implementing the pre-big bang curvaton model for the axion
\cite{Enqvist:2001, Sloth:2002}, but keeping the dilaton 
perturbations would lead to our spectrum \re{spectrum2} with
the spectral indices $n_1=4$ and $n_2=1$, which incidentally
agree with our best-fit model.

\begin{table}[!tb]
\small
\begin{tabular}{| l | r r r r r |} \hline
                  				& Adiabatic\hspace{\fill} & Our\hspace{\fill} & Our\hspace{\fill}  &  Adiabatic & Our model \\ 
                  				& best-fit             & best-fit     & example     &                            &           \\ \hline
 $\chi^2$         				& 1429.0               & 1423.9       & 1425.9      &                            &                      \\
 $\nu$            				& 1342                 & 1337         & 1337        &                            &                      \\
 $\chi^2/\nu$     				& 1.065                & 1.065        & 1.066       &                            &                      \\ \hline
 Primary parameters  				&                      &              &             &                            &                      \\
 $\omega_b$ \hspace{\fill} [0.005, 0.1]		& 0.0229 	       & 0.0409       & 0.0284      & \m{0.0238}{0.0015}{0.0026} & \m{0.0315}{0.0044}{0.0107} \\
 $\omega_{dm}$ \hspace{\fill} [0.01, 0.99]	& 0.122                & 0.115        & 0.150       & \m{ 0.118}{ 0.018}{ 0.018} & \m{ 0.125}{ 0.024}{ 0.023} \\
 $100\times\Theta$\hspace{\fill} [0.3, 10] 	& 1.044                & 1.074        & 1.060       & \m{1.049}{0.007}{0.008}    & \m{ 1.062}{ 0.013}{ 0.016} \\
 $\tau$ \hspace{\fill} [0.01, 0.8]   		& 0.115                & 0.498        & 0.127       & \m{ 0.159}{ 0.074}{ 0.157} & \m{ 0.222}{ 0.114}{ 0.227} \\
 $f_\nu$ \hspace{\fill} [0, 1] 			& ---                  & 0.855        & 0.725       & ---                        & $>$ 0.8 \\
 $n_1$ \hspace{\fill} [0.2, 4]     		& 0.964                & 3.906        & 1.065       & \m{ 0.992}{ 0.037}{ 0.083} & \m{ 1.377}{ 0.289}{ 1.600} \\
 $\Delta n$ \hspace{\fill} [0, 3.8]  		& ---                  & 2.918        & 0.220       & ---                        & ($<$ 1.06) \\
 $\ln(10^{10} N_\sR^2)$  \hspace{\fill} [1, 6] 	& 3.950                & 4.492        & 3.884       & \m{ 4.054}{ 0.166}{ 0.310} & \m{ 4.031}{ 0.242}{ 0.413} \\
 $\cos\phi$ \hspace{\fill} [0, 1]       	& (1)                  & 0.295        & 0.422       & (1)                        & no constraint \\
 $\cos\theta$ \hspace{\fill} [-1, 1]  		& ---                  & 0.985        &  0.731      & ---                        & $>$ 0.76 \\
 $f_\sS$ \hspace{\fill} [-42, 42]      		& (0)                  & -9.156       & -1.400      & (0)                        & $|f_\sS|<$ 13 \\ \hline
 Derived parameters  				&              	       &              &             &                            &                \\ 
 $n_2$      					& ---     	       & 0.988        & 0.845	    & ---                        & \m{ 0.869}{ 0.160}{ 0.205} \\
 $\tilde f_\sS = (1-f_\nu)f_\sS$  		& (0) 		       & -1.329       & -0.385      & (0)                        & $|\tilde f_\sS|<$ 0.7 \\
 $\Omega_\Lambda$                               & 0.695                	& 0.736       & 0.345       & \m{ 0.726}{ 0.084}{ 0.087} & \m{ 0.627}{ 0.243}{ 0.141} \\
 $\Omega_b$ 					& 0.048  	       & 0.069        & 0.103 	    & \m{ 0.046}{ 0.008}{ 0.008} & \m{ 0.079}{ 0.019}{ 0.025} \\
 $\Omega_m$                                     &  0.305  	       & 0.264        & 0.655       & \m{ 0.274}{ 0.087}{ 0.084} & \m{ 0.373}{ 0.141}{ 0.244} \\
 $\Omega_{cdm}$                                 &  0.257  	       & 0.028        & 0.151       & \m{ 0.229}{ 0.079}{ 0.076} & $<$ 0.06 \\
 $\Omega_\nu$                                   &  ---    	       & 0.166        & 0.399       & ---    			 & \m{ 0.239}{ 0.124}{ 0.213} \\
 Age (Gyr)                                      & 13.62                & 12.68        & 14.32       & \m{13.39}{ 0.51}{ 0.30}    & \m{13.69}{ 1.16}{ 0.95} \\
 $z_{re}$                                       & 13.73                & 25.93        & 13.45       & \m{16.86}{ 5.73}{ 7.78}    & \m{18.41}{ 7.78}{ 7.87} \\
 $H_0$ (km/s/Mpc)                               & 68.87                & 77.02        & 52.21       & \m{71.78}{ 5.36}{10.10}    & \m{64.29}{11.38}{15.87}  \\ \hline
 Low-$l$ behaviour                              &                      &              &             & \hspace{\fill}     & \hspace{\fill}   \\
 $2(2+1)C_2/2\pi$ ($\mu$K$^2$)                  & 1170    	       & 1045         & 736         & (mean) \m{1262}{190}{202}         &  (mean) \m{1081}{226}{225}\\
 $\chi^2(C_2,C_3)$                              &  7.338               & 6.408        & 4.336       & (mean) 7.888               & (mean) 6.615   \\
 $\chi^2(C_2,C_3)/\chi^2$ (\%)                  & 0.514   	       & 0.450        & 0.304       & (mean) 0.550               & (mean) 0.461   \\ \hline
\end{tabular}
\caption{Comparison of models.
In the last two columns we give the median of the
1-dimensional marginalised likelihood for each parameter
and the region around the median that contains
68\% of the accepted models in our chains.
If the likelihood function was exactly Gaussian, this would
correspond to the $1\sigma$ region. However, most of
the likelihood functions are too far from Gaussian
for this to be true, as should be clear from
the values of our best-fit model. Thus the parameter ranges
should not be taken as $1\sigma$ regions.
They are given {\it for comparison purposes only.}
}
\label{table:parameters}
\end{table}

\subsection{Likelihood distributions} \label{likelihood}

Let us briefly discuss the main features
of the marginalised likelihood distributions,
concentrating on the differences between
our isocurvature model and the pure adiabatic
$\Lambda$CDM model.

In Table \ref{table:parameters} we
give the median values for the primary and derived parameters,
and the regions that contain 68\% of the accepted models in
the 1-dimensional marginalised likelihood distributions for
the adiabatic model and for our model. As the likelihood
function is non-Gaussian with respect to some
parameters, most of the quoted numbers cannot be
interpreted as 1$\sigma$ confidence levels and are
for comparison purposes only.
(The meaning of the given values is as described
in the appendix of \cite{Tegmark:2003ud}.) Of the primary
and derived parameters, only $\omega_{dm}$, $\Theta$, $n_2$,
$\Omega_{cdm}$ and the age of the universe have a strongly Gaussian 1-d
marginalised likelihood distribution. Of course, even
for these parameters, the numbers are confidence ranges,
and should not be treated as exclusion limits, as emphasised
in \cite{Trotta:2002}.
Adding more input into our analysis, such
as the SDSS data \cite{Tegmark:2003ud, Tegmark:2003uf}
would lead to better behaved likelihood distributions, in
agreement with the discussion above.

The well-known degeneracy between the physical
baryon density $\omega_b$ and isocurvature modes
\cite{Valiviita:2003a, Bucher:2004, Trotta:2001}
is apparent also from our results. In a pure adiabatic power-law
model, $\omega_b$ is determined simply by the 
relative heights of the successive acoustic peaks.
Since the CDM isocurvature mode has valleys
where the adiabatic mode has peaks, increasing the
relative isocurvature contribution can compensate for
adding baryons.
This, and the fact that the modes are the sum of
two power-laws, means that there is much more freedom
for the values of $\omega_b$ in our model than in a pure
adiabatic model. Also, the preferred values of $\omega_b$
with isocurvature modes are generally higher than in the
pure adiabatic case.

The optical depth $\tau$ is poorly constrained in
the pure adiabatic case, and even more so in our model.
The 95\% region spans a huge range
from 0.05 to 0.59, and the 68\% region given in
Table \ref{table:parameters} is also wide. The
optical depth is primarily determined by the single data
point at $l=2$ in the WMAP TE data. This quadrupole TE power
is very high, hinting at early reionisation
(large optical depth) in the pure adiabatic case.
In our model the mechanism that suppresses the
low TT multipoles also cancels TE power,
which means that an even larger $\tau$ is needed
to fit the high first data point; on the other hand,
positive correlation would add TE power at the quadrupole
and allow a smaller $\tau$. Note that in the pure adiabatic
case, the signal for the large optical depth 
in the TT spectrum comes completely from the southern
hemisphere; the preferred value for the northern hemisphere is
$\tau=0$ \cite{Hansen:2004c}. Such asymmetry suggests caution
about the interpretation of the data points leading to the value of
$\tau$. At any rate, the effect on our results is small, since
they are consistent with a wide range of values.

In the pure adiabatic power-law model there is only one
spectral index while our model has two, so a simple comparison
of their values is not very meaningful. However,
let us mention that while the peak of the likelihood function
for the larger spectral index $n_1$ is at the moderate value
1.1, the distribution has a long tail at large values, so that
the median is 1.4 and the mean is even larger, 1.8.
While this may be an artifact of poor convergence of
$\Delta n$ (the split test shows that the upper bound for
$\Delta n$ is not reliable), our best-fit model with $n_1=3.9$
demonstrates that the WMAP data does not require the spectrum
to be almost scale-invariant over the entire range of scales
covered by the data.

The vacuum energy density $\Omega_\Lambda$ is
much less constrained than in pure adiabatic models. The
reason is again low-$l$ modifications. Increasing
$\Omega_\Lambda$ raises the TT power at the lowest
multipoles due to the integrated Sachs-Wolfe effect. This
effect can be countered by negative correlation
which brings the low multipoles down, or mimicked
by positive correlation which takes them up. The
net effect is to widen the range of $\Omega_\Lambda$
in both directions. The Hubble parameter and vacuum
energy density are strongly correlated, so the freedom in the
values of $\Omega_\Lambda$ is reflected in the values of $H_0$.

It is notable that our isocurvature spectrum
prefers a large neutrino fraction $f_\nu$, in contrast
to a pure adiabatic spectrum which is not very sensitive
to the value of $f_\nu$
\cite{Tegmark:2003ud, Tegmark:movie, Elgaroy:2003, Hannestad:2004nb}.
Our model differentiates between CDM and massive neutrinos
because the latter carry no isocurvature perturbations. This
means that for fixed dark matter density and CDM isocurvature
amplitude, one can tune down the isocurvature contribution
by increasing the neutrino fraction $f_\nu$: the less CDM there
is, the less difference its perturbations make. The 
total dark matter-photon isocurvature perturbation $\sS_{dm,\gamma}$
and cold dark matter-photon isocurvature
perturbation $\sS_{cdm,\gamma}$ are related by
$\sS_{dm,\gamma}=(1-f_\nu)\sS_{cdm,\gamma}$, so
the power spectrum is almost (but not entirely)
degenerate with respect to scaling $f_\sS$ and $1/(1-f_\nu)$
by the same factor. As it is, our model seems to prefer
a high CDM-photon isocurvature fraction $f_\sS$, with
a large neutrino fraction to compensate. 
A meaningful measure of the isocurvature contribution
is given by the dark matter-photon isocurvature fraction
$\tilde f_\sS = (1-f_\nu) f_\sS$. We find
$|\tilde f_\sS| < 0.7$ in the 68\% region (in agreement with
\cite{Valiviita:2003a, Peiris:2003, Crotty:2003}), though a
look at our best-fit model with $\tilde f_\sS=-1.3$ should
again serve as a warning not to interpret this number as an
exclusion limit.

Since $f_\sS$ and $1-f_\nu$ are almost
degenerate, measuring one will constrain the other.
As CDM and neutrinos
have a very different effect on the matter power spectrum
\cite{Tegmark:2003ud, Tegmark:movie},
using the SDSS data to give a handle on the neutrino fraction
could indirectly constrain the isocurvature
contribution (which is itself negligible at the SDSS scales
since the isocurvature perturbations are damped rapidly
with increasing $l$).

Finally, from Table \ref{table:parameters} we see that
the mean TT quadrupole power in the MC chains for our model is
181 $\mu$K$^2$ smaller than for pure adiabatic models. This shows
that the cancellation mechanism works
and on average helps in fitting the low-$l$
part of the spectrum. The mean $\chi^2$ contribution of the first two
TT data  points to the total $\chi^2$ is also reduced from 0.55\%
to 0.46\%. We devote the next section to this important issue.

\subsection{Suppression of the low multipoles} \label{suppression}

The most important feature of the angular
power spectrum that distinguishes the inflaton-curvaton
model from the standard adiabatic power-law $\Lambda$CDM
model is the amplitude of the low multipoles.
In fact, this is the only significant qualitative
difference, which is not surprising. As discussed in section
\ref{anticorrelation}, the pure adiabatic $\Lambda$CDM model
fits the TT power spectrum very well, apart from the low
multipoles and the three ``blips'' at intermediate multipoles
before and at the first acoustic peak (see Figure
\ref{figure:clcomp}). The three blips cannot be fitted with
the smooth CDM isocurvature modes, so the
only place where we can get significant improvement is
the low multipoles.

While the discrepancy of the low multipoles
in the adiabatic $\Lambda$CDM model looks significant
to the eye (see Figure \ref{figure:clcomp}), the small
number of badly fitted data points (practically the first two points)
combined with the large cosmic variance means that the weight
of this feature in the overall $\chi^2$ fit is quite small
(for the best-fit adiabatic model only 0.51\%, or 7 points, of the
total $\chi^2$).
Therefore one does not expect any model that does not explain
the ``blips'' to have a significantly lower $\chi^2$ than the pure
adiabatic $\Lambda$CDM case, regardless of what the model is.
Indeed, the difference in $\chi^2$ between the best-fit pure
adiabatic model and the best-fit inflaton-curvaton model is
only 5: our model fits better, but not decisively better.

However, the small amplitude of the low multipoles is an
interesting feature beyond its statistical significance,
and has attracted a lot of attention. Since a $\chi^2$-selection
among models does not particularly pick out ones with small
amplitude for the low multipoles (in our best-fit model,
less than 1 point of the improvement in the $\chi^2$ comes
from the quadrupole and octopole), we specifically look
for low quadrupole power among those of our models that
fit the data nearly as well as the best-fit model.
(The inclusion of data from high-$l$ CMB
measurements and SDSS would further reduce the weight of the
low multipoles in the overall fit, making it more
difficult to find models with a low quadrupole.)

\begin{figure*}[p]
  \includegraphics[angle=270,width=0.88\textwidth]{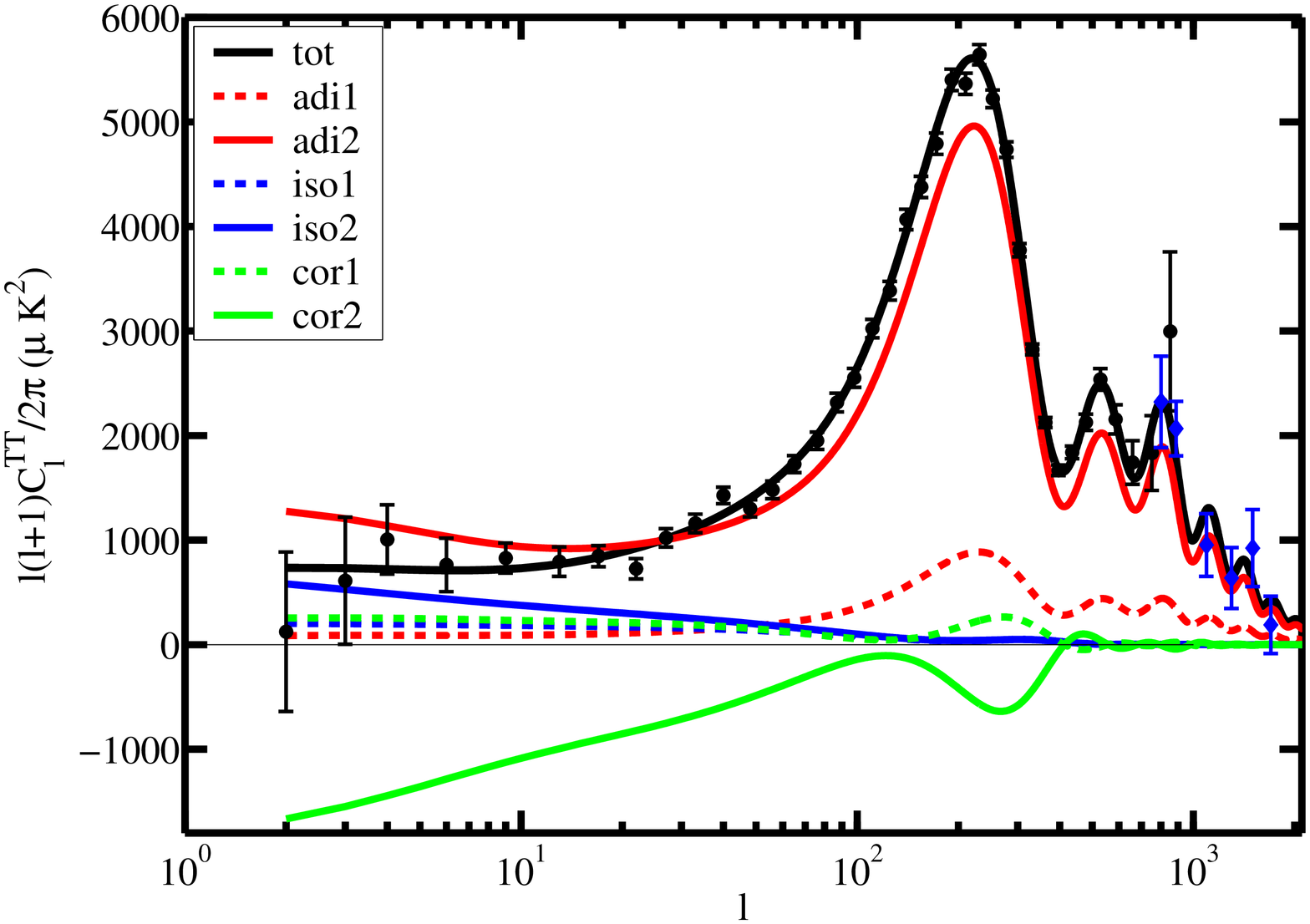}\\ 
  $\phantom{X}$ \includegraphics[angle=270,width=0.9\textwidth]{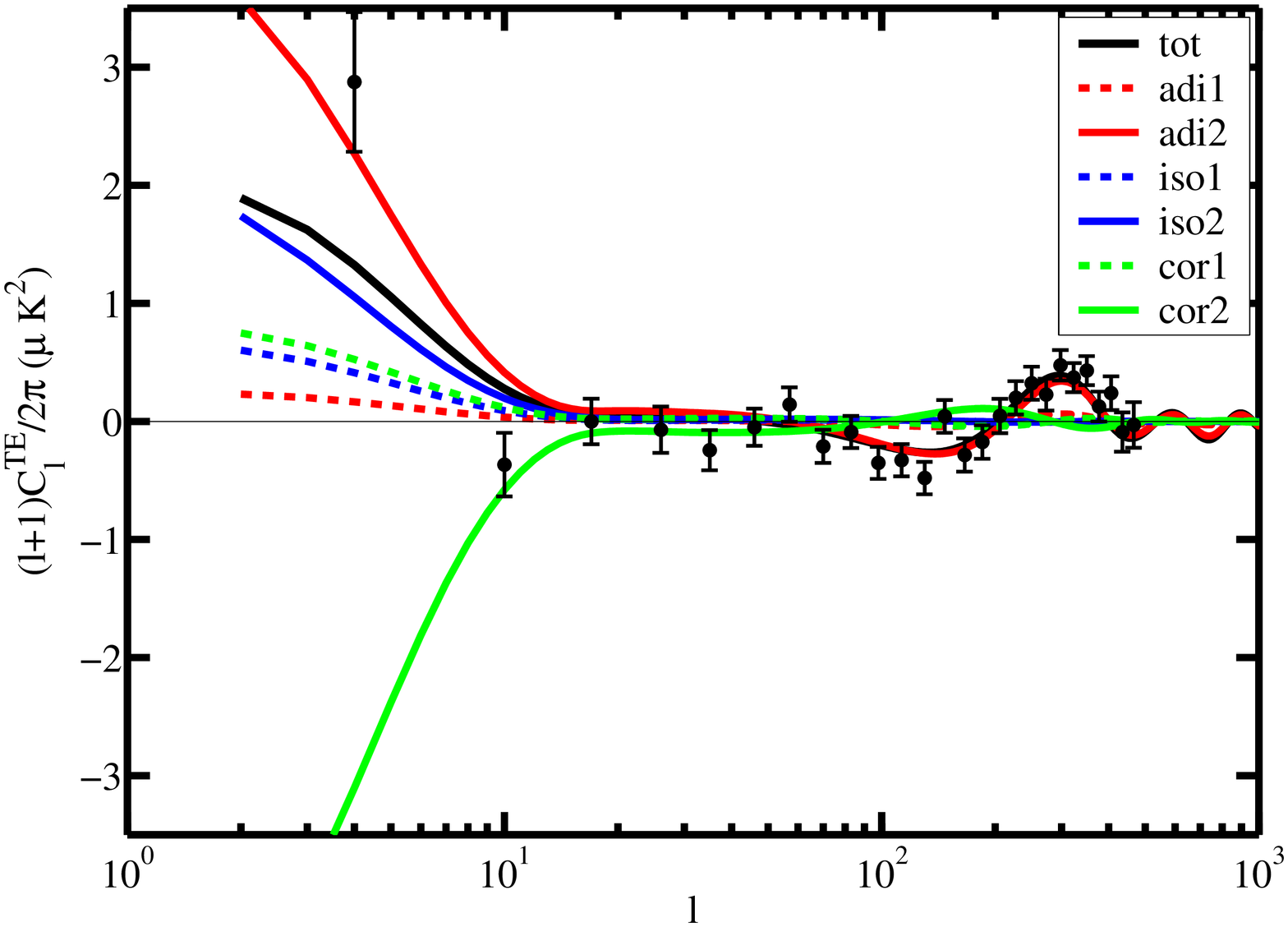}
  \caption{The example model of the third column of Table
    \ref{table:parameters} with the WMAP data ($\bullet$) and
    the highest $l$ data points of other CMB experiments from Tegmark's
    compilation \cite{Tegmark:movie} (\textcolor{blue}{$\blacklozenge$}).
    The total angular power consists of the six components
    mentioned in the paragraph after equation
    \re{ClTE}: $C_l^{\rm tot} = C_l^{\rm adi1} +
    C_l^{\rm iso1} + C_l^{\rm cor1} + C_l^{\rm adi2} +
    C_l^{\rm iso2} + C_l^{\rm cor2}$. Upper panel: Temperature-temperature
    angular power $l(l+1)C_l^{TT}/2\pi$. Lower panel: Temperature-polarisation    cross-correlation power $(l+1)C_l^{TE}/2\pi$.}
 \label{figure:cl}
\end{figure*}

\begin{figure*}[t]
  \includegraphics[angle=270,width=0.9\textwidth]{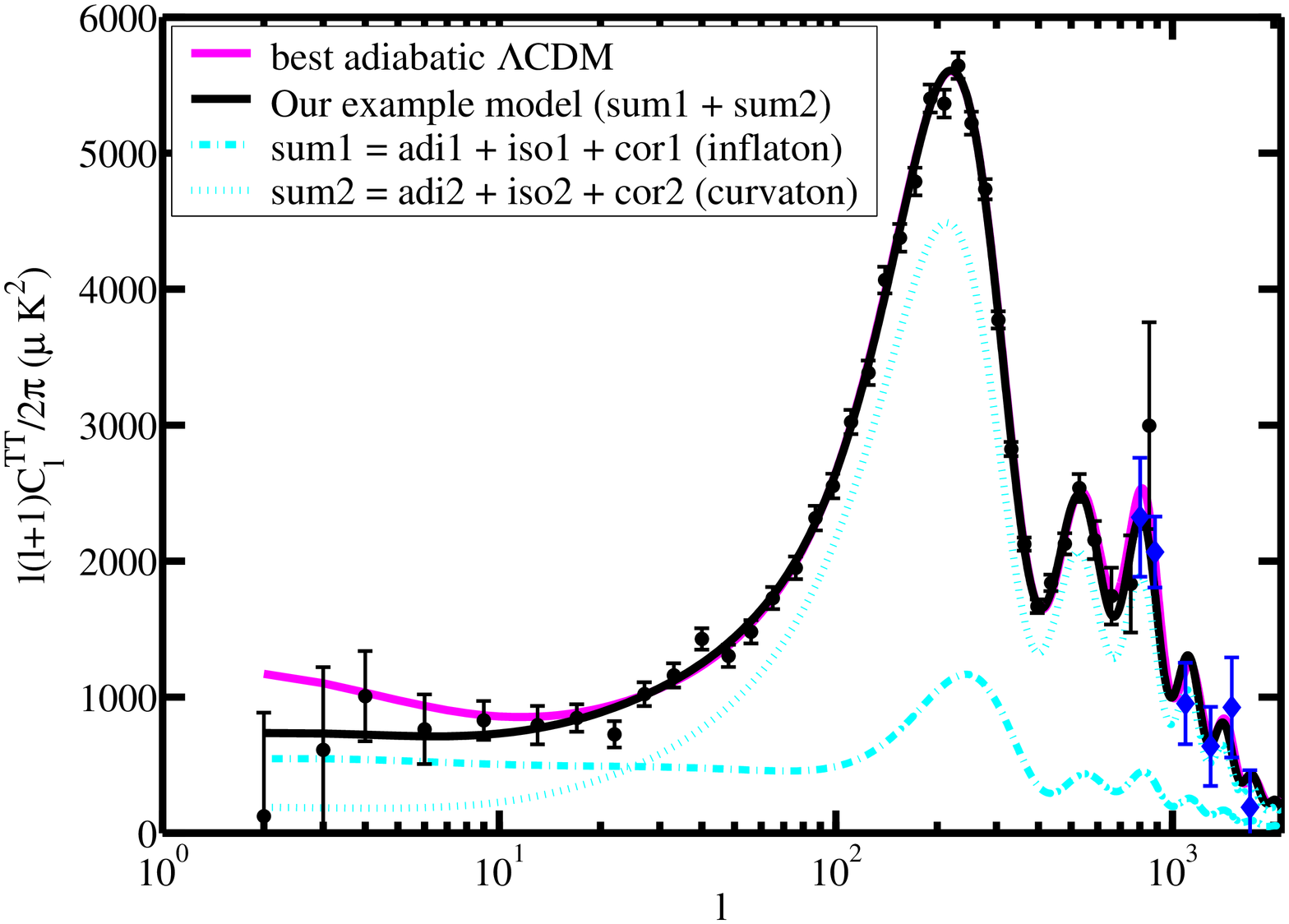}
  \caption{Comparison of our example model
  (the third column of Table \ref{table:parameters}, see also
  Figure \ref{figure:cl}) to the best-fit 6-parameter adiabatic $\Lambda$CDM
  model (the first column of Table \ref{table:parameters}).
  The total angular power of our model is the black solid line;
  $C_l^{\rm tot} = C_l^{\rm sum1} + C_l^{\rm sum2}$, where
  $C_l^{\rm sum1} = C_l^{\rm adi1} + C_l^{\rm iso1} + C_l^{\rm cor1}$
  and 
  $C_l^{\rm sum2} = C_l^{\rm adi2} + C_l^{\rm iso2} + C_l^{\rm cor2}$.}
\label{figure:clcomp}
\end{figure*}

We take a model found using this
method as our example with which to demonstrate the
suppression mechanism of the low multipoles.
The parameters for the example model are given in 
the third column of Table \ref{table:parameters}.
The TT and TE angular power spectra of the example model
and the contributions from the individual adiabatic,
isocurvature and correlation components
are shown in Figure \ref{figure:cl}.
In Figure \ref{figure:clcomp} the inflaton and curvaton
contributions and the total TT spectrum are plotted in
comparison to the adiabatic power-law $\Lambda$CDM model.

The example model has $\chi^2 = 1425.9$, which is 2 larger than
our best-fit, and still 3 smaller than the adiabatic best-fit.
On the last three lines of
Table \ref{table:parameters} we compare the best-fit
adiabatic $\Lambda$CDM model, our best-fit model and our
example model.
We give the quadrupole power, the $\chi^2$ of the first
two data points (quadrupole and octopole) and their
contribution to the total $\chi^2$. The quadrupole
power in our example model is 736 $\mu$K$^2$, which is
434 $\mu$K$^2$ smaller than in the best-fit adiabatic
model. This makes the fit better on
the SW plateau and drops the $\chi^2$ contribution of the
first two data points from 0.51\% to 0.30\%. Note that
unlike in our best-fit model, practically all of the
improvement in the $\chi^2$ over the adiabatic case comes
from the low multipoles.

For comparison purposes, we also did a search for
low quadrupole power among $\Lambda$CDM models that
are within $\Delta\chi^2 < 2$ from the best-fit $\Lambda$CDM model.
As expected, the lowest quadrupole power found (914 $\mu$K$^2$) 
is significantly higher than in our example model.

\paragraph{Extreme cancellation.}

Let us look at the details of the suppression mechanism of the
low multipoles with the spectrum \re{spectrum2}.
If we wanted to pull the amplitude on the SW plateau
down as much as possible, we could put $\cos\phi=0$ in \re{Cla}
to kill $C_l^{\rm adi1}$ and $C_l^{\rm cor1}$, and then use
anticorrelation to get rid of
$C_l^{\rm adi2}+C_l^{\rm iso2}+C_l^{\rm cor2}$ on the SW plateau.
Given that $g_{l,\sS}^T(k)=-2 (1-f_{\nu}) g_{l,\sR}^T(k)$
on the SW plateau, this implies, from \re{Clb}, that
$\sin\phi + 2 (1-f_{\nu}) f_{\sS} \sin\theta=0$.
The TT spectrum would then be
\bea \label{ClSW}
  \fl C_l^{TT} = 4\pi \int\frac{\rmd k}{k} N_{\sR}^2 \left[ \left(\frac{g_{l,\sS}^T(k)}{2 (1-f_{\nu})} \right)^2 \cot^2\theta\, \kt^{n_1-1} + \left( g_{l,\sR}^T(k) + \frac{g_{l,\sS}^T(k)}{2 (1-f_{\nu})} \right)^2 \kt^{n_2-1} \right] \ ,
\eea

\noindent and we have $\cos\phi=0$, $\tilde f_{\sS}=-1/(2 \sin\theta)$.
On the SW plateau, the second term is zero, and the spectrum
has the index $n_1$, with an amplitude that is freely adjustable
with $\cot^2\theta$. Away from the plateau, the isocurvature
transfer function $g_{l,\sS}^T(k)$ is damped more rapidly than the
adiabatic transfer function $g_{l,\sR}^T(k)$, so the spectrum at
high multipoles is essentially adiabatic, has the spectral
index $n_2$ and an amplitude given by $N_{\sR}^2$.
Since we have $n_1>n_2$, the low multipoles are further
suppressed relative to the high multipoles.

In other words, the low multipoles are given by
inflaton isocurvature perturbations with the larger
spectral index $n_1$, while the peak structure is given
by adiabatic curvaton perturbations with the smaller
spectral index $n_2$. The SW plateau and the peak structure 
can therefore be adjusted almost independently to
fit the data.

\paragraph{Cancellation in our example model.}

In contrast to the extreme case, in our
example model the behaviour of the high multipoles is
not completely decoupled from that of the low multipoles.
However, the cancellation mechanism for the low multipoles
works qualitatively as sketched above: a small $\cos\phi$
suppresses the adiabatic and correlation components
$C_l^{\rm adi1}$ and $C_l^{\rm cor1}$ with
the larger spectral index $n_1$, and strong
anticorrelation suppresses all three components
$C_l^{\rm adi2}$, $C_l^{\rm iso2}$ and $C_l^{\rm cor2}$
with the smaller spectral index $n_2$.

From Figures \ref{figure:cl} and \ref{figure:clcomp}
we see that the curvaton perturbations with the
spectral index $n_2=0.85$ are killed
almost completely by the anticorrelation at the
low multipoles, but they dominate the peak structure.
The power at low multipoles comes almost completely
from the inflaton perturbations, with $n_1=1.07$. However,
in contrast to the extreme case, the amplitude of the
adiabatic component $C_l^{\rm adi1}$ is not exactly zero,
and it gives a significant subdominant contribution
to the acoustic peaks. For the first peak, the inflaton
and curvaton correlation modes $C_l^{\rm cor1}$ and
$C_l^{\rm cor2}$ also contribute. The correlation results
in peak structure between pure adiabatic and 
pure isocurvature cases, and the fact that those
components are themselves a sum of two power-laws
also slightly modifies the relative heights
of the successive peaks (and valleys).

Since our example model has a non-zero vacuum energy
density ($\Omega_\Lambda = 0.35$), the integrated SW effect
would typically lead to rising power towards the smallest
multipoles (as apparent for the adiabatic $\Lambda$CDM model
in Figure \ref{figure:clcomp}). However, the inflaton
perturbations are going down with increasing $l$
because they are dominated by the isocurvature component while
the curvaton perturbations are slowly rising with increasing $l$
as the cancellation by anticorrelation weakens after $l\sim10$
(as the SW effect ceases to be dominant), and this
interplay keeps the spectrum quite flat at low multipoles.

Not all features of our example model are universal
to well fitting models with the spectrum \re{spectrum2}.
As discussed in section \ref{fitresults}, good
models do not cluster around a single region in parameter space.
However, the suppression mechanism which combines
anticorrelation and two different spectral indices is indeed a
generic feature of our spectrum \re{spectrum2},
and is not tied to the exotic features (such as the high
physical baryon density $\omega_b$) of our example model.

\paragraph{Running of the spectral index.}

Analysis of CMB data with pure adiabatic models with
a running spectral index has pointed towards
the feature that the spectral index evolves
from a large to a small value as $k$ increases
\cite{Peiris:2003, Bridle:2003, Bond:2004rt}, though the evidence is marginal.
(Note also \cite{Blanchard:2003} where the fit to the data is
improved by having an index that does not run but changes
discontinuously from a large to a small value.)

For a pure adiabatic spectrum which is a sum of two
different power-laws, the effective running of the spectral
index is always positive: the smaller index dominates at
small $k$ and the larger index at large $k$. As discussed
above, in our model it is natural for the large index to
dominate at small $k$ and vice versa.
Of course, suppression of the low multipoles via
anticorrelation itself also looks like negative
running of the spectral index if one tries to fit
it with a pure adiabatic power-law. Indeed,
the indications of running in the WMAP data come mostly
from the amplitude of the low multipoles \cite{Bridle:2003}.

Correlated isocurvature perturbations
allow negative running because the prefactors evolve
with $k$, as seen in \re{Clb}. At small $k$, the prefactor of the
perturbations with the small index is suppressed by
anticorrelation, while at large $k$ the prefactor of the
perturbations with the large index is suppressed by the rapid
damping of isocurvature perturbations with $k$.
A key feature allowing the complete cancellation
of the adiabatic terms at the low multipoles and thus the
decoupling of the low and high multipoles
is that the adiabatic and isocurvature spectra have
the same shape, a combination of two power-laws.

Previous studies where the adiabatic and isocurvature spectra
have been taken to obey a simple power-law with the same index
\cite{Gordon:2002, Langlois:1999b, Moroi:2003} have found similar
suppression. However, that form of the spectrum is too constrained
for the low and high multipoles to be decoupled.
For more complicated spectra, the adiabatic and
isocurvature components have often not had the same shape
\cite{Valiviita:2003a, Amendola:2001, Peiris:2003, Valiviita:2003b, Crotty:2003},
so they cannot effectively cancel, and the low multipoles
again cannot be as decoupled from the high ones
and as effectively suppressed as in our case. (However, double
inflation models with spectra similar to \re{spectrum2}
have been studied \cite{Langlois:1999a, Tsujikawa:2002}.)
In the case of models with more than one isocurvature
component, but all having the same spectral index
\cite{Bucher:2004, Trotta:2001, Trotta:2002}, the fit to the
data has also improved less over the adiabatic case than
for our spectrum \re{spectrum2}. Note that this is not
a question of the number of parameters: for example,
the models studied in \cite{Valiviita:2003a, Bucher:2004}
have as many or more parameters than our model.
This underlines the importance of priors in 
the form of the spectrum on the estimation of the
isocurvature contribution.

\section{Conclusion}

We have studied a model that generalises the usual curvaton
scenario in two ways. First, we model the curvaton
properly as a scalar field instead of dust when it decays.
We follow the curvaton decay numerically. We find that
the results of the dust case are recovered whenever the curvaton
does not significantly contribute to the energy density when it
is starting to oscillate, even if the curvaton does not
oscillate rapidly when it decays.

Second, we take into account that in addition to the curvaton,
the inflaton can also have perturbations. The resulting
spectrum is an interesting mixture of correlated adiabatic
and CDM isocurvature perturbations, which could arise also
from a double inflation model. We analyse this spectrum
using the first year WMAP data. We find that including the
isocurvature modes opens up the parameter space of well
fitting models,
and the parameters for both the perturbations and the
cosmological background no longer cluster around the best-fit
of the standard adiabatic power-law $\Lambda$CDM model.
For example, the spectral indices do not
need to be close to scale-invariant, and the physical baryon
density $\omega_b$ is typically much larger than the usual value.
Adding other input, such as data from high-$l$
CMB measurements and SDSS would exclude many of the new
well-fitting models. We will consider these issues in a
more thorough analysis of the perturbation spectrum in a
follow-up paper \cite{Ferrer:2004}.

The most significant qualitative feature of our model
is the suppression of the low multipoles due to
anticorrelation between adiabatic and isocurvature
perturbations. Our spectrum makes it possible
to decouple the behaviour of the low and high multipoles
and thus fit the observed small amplitude of the quadrupole
and octopole without affecting the adiabatic peak structure,
improving the fit to the data. There is no need to fine
tune the multipoles affected by the suppression: if
the perturbations in the decay products of two (or more) scalar
fields both contribute to the radiation and matter observed
today, it is the low multipoles that are generically affected by
the resulting isocurvature contribution.
Recently, another model where the suppressed multipoles
do not have to be picked by hand, but are instead related to
the dark energy via an IR/UV duality, has been proposed \cite{Enqvist:2004xv}.

The significance of having qualitatively different
behaviour of the spectrum at small and large $k$ goes beyond
explaining the small amplitude of the low multipoles, because the
determination of cosmological parameters is sensitive to the
behaviour at small $k$. For example, in \cite{Blanchard:2003}
a model with pure adiabatic perturbations with a different
spectral index for small and large $k$ and no cosmological constant
was shown to fit the CMB and large scale structure data
better than the $\Lambda$CDM model.

There are indications that the WMAP signal for the quadrupole
as well as other low multipoles, up to $l=40$,
has significant non-cosmological contamination
\cite{Hansen:2004c, Eriksen:2003, Hansen:2004a, Schwarz:2004, Hansen:2004b, Slosar:2004}.
This could completely change the estimation of isocurvature
perturbations, since their contribution is determined mostly
from the low multipole part of the spectrum. On the other
hand, the presence of such contamination would underline the
importance of studying how cosmological parameter estimation
is affected by possibly obscured features at small $k$,
such as isocurvature modes.

\ack

SR thanks Subir Sarkar for helpful discussions and
comments on the manuscript, and
William H. Kinney for correspondence.
JV thanks Hannu Kurki-Suonio for discussions.
FF has been supported by the Leverhulme trust,
SR by PPARC grant PPA/G/O/2002/00479 and by the
European Union network HPRN-CT-2000-00152,
``Supersymmetry and the Early Universe'', 
JV by the Magnus Ehrnrooth foundation 
and by the Research Foundation of the University of Helsinki
(Grant for Young and Talented Researchers).
We acknowledge CSC -- Scientific Computing Ltd (Finland) --
for computational resources.\\

\end{document}